%% file: narcissus.tex
\documentclass[acmsmall]{acmart}
\renewcommand\footnotetextcopyrightpermission[1]{} 
\acmJournal{PACMPL}
\acmVolume{1}
\acmNumber{CONF} 
\acmArticle{1}
\acmYear{2018}
\acmMonth{1}
\acmDOI{} 
\startPage{1}

\setcopyright{none}

\usepackage{booktabs}   
\usepackage{subcaption} 
\usepackage{amsmath, amssymb}
\usepackage{hyperref}           
\usepackage{sfmath}             
\usepackage{tikz}               
\usepackage{listings}           
\usepackage{microtype}          
\usepackage{bcprules}           
\usepackage{fancyvrb}           
\usepackage{wrapfig}            
\usepackage{graphicx}           
\usepackage{hhline}             
\usepackage{lipsum}             
\usepackage{mathtools}          
\usepackage{mdframed}             
\usepackage{algorithm}          
\usepackage{varwidth}
\usepackage[noend]{algpseudocode}

\usepackage{enumitem}           

\usetikzlibrary{shapes.geometric} 

\algrenewcommand\algorithmicindent{1.0em}%
\algrenewcommand\textproc[1]{\textsf{#1}}
\algnewcommand\algorithmicmatch{\textbf{match}}
\algnewcommand\algorithmiccase{\textbf{case}}
\algdef{SE}[MATCH]{Match}{EndMatch}[1]{\algorithmicmatch\ #1\ \textbf{with}}{\algorithmicend\ \algorithmicmatch}%
\algdef{SE}[CASE]{Case}{EndCase}[1]{\algorithmiccase\ #1  $\Rightarrow$}{\algorithmicend\ \algorithmiccase}%
\algdef{SE}[Try]{Try}{EndTry}{\textbf{try}}{\algorithmicend\ \textbf{try}}%
\algtext*{EndMatch}%
\algtext*{EndCase}%
\algtext*{EndTry}%

\input{Preamble}

\begin{document}

\title[Narcissus]
{\textsc{Narcissus}: Correct-By-Construction Derivation of
  Decoders and Encoders from Binary Formats}


\author{Benjamin Delaware}
\affiliation{\institution{Purdue University}}            
\email{bendy@purdue.edu}          

\author{Sorawit Suriyakarn}
\affiliation{\institution{Hudson River Trading}}            
\email{sorawit@csail.mit.edu}          

\author{Clément Pit-Claudel}
\affiliation{\institution{MIT CSAIL}}            
\email{cpitcla@csail.mit.edu}          

\author{Qianchuan Ye}
\affiliation{\institution{Purdue University}}            
\email{ye202@purdue.edu}          

\author{Adam Chlipala}
\affiliation{\institution{MIT CSAIL}}            
\email{adamc@csail.mit.edu}          

\begin{abstract}
  It is a neat result from functional programming that libraries of \emph{parser combinators} can support rapid construction of decoders for quite a range of formats.
  With a little more work, the same combinator program can denote both a decoder and an encoder.
  Unfortunately, the real world is full of gnarly formats, as with the packet formats that make up the standard Internet protocol stack.
  Most past parser-combinator approaches cannot handle these formats, and the few exceptions require redundancy -- one part of the natural grammar needs to be hand-translated into hints in multiple parts of a parser program. %
  We show how to recover very natural and nonredundant format specifications, covering all popular network packet formats and generating both decoders and encoders automatically.
  The catch is that we use the Coq proof assistant to derive both kinds of artifacts using tactics, automatically, in a way that guarantees that they form inverses of each other.
  We used our approach to reimplement packet processing for a full Internet protocol stack, inserting our replacement into the OCaml-based MirageOS unikernel, resulting in minimal performance degradation.
\end{abstract}


\widowpenalty=50
\clubpenalty=50
\brokenpenalty=50
\displaywidowpenalty=50

\maketitle

\section{Introduction}
\label{sec:Introduction}

Decoders and encoders are vital components of any software that
communicates with the outside world, and accordingly functions that
process untrusted data represent a key attack surface for malicious
actors.  Failures to produce or interpret standard formats routinely
result in data loss, privacy violations, and service outages in
deployed systems~\cite{CVE-2015-0618, CVE-2013-1203,
  CVE-2012-5965}. In the case of formally verified systems, bugs in
encoder and decoder functions that live in the unverified, trusted
code have been shown to invalidate the entire assurance
case~\cite{Fonseca+17}. There are no shortage of code-generation
frameworks~\cite{YACC79, ANTLR1995, XDR, ASN2001, Avro, protobuf,
  Binpac2006, PADS2005, PacketTypes, Datascript, Nail, Next700} that
aim to reduce opportunities for user error in writing encoders and
decoders, but these systems are quite tricky to get right and have
themselves been sources of serious security bugs~\cite{CVE-2016-5080}.

\emph{Combinator libraries} are an alternative approach to the rapid
development of parsers which has proven particularly popular in the
functional-programming community~\cite{leijen+2001}. This approach has
been adapted to generate both parsers and pretty printers from single
programs~\cite{Kennedy+2004, Rendel+2010}. Unfortunately, combinator
libraries suffer from the same potential for bugs as code-generation
frameworks, with the additional possibility for users to introduce
errors when extending the library with new combinators. This paper
presents \frameworkName, a combinator-style framework for the Coq
proof assistant that eliminates the possibility of such bugs, enabling
the derivation of encoders and decoders that are correct by
construction. Each derived encoder and decoder is backed by a
machine-checked functional-correctness proof, and \frameworkName
leverages Coq's proof automation to help automate both the
construction of encoders and decoders and their correctness
proofs. Key to our approach is how it threads information through a
derivation, in order to generate decoders and encoders for the sorts
of non-context-free languages that often appear in standard networking
protocols.

We begin by introducing the key features of \frameworkName
with a series of increasingly complex examples, leading to a
hypothetical format of packets sent by a temperature sensor to a smart
home controller. In order to build up the reader's intuition, we
deliberately delay a discussion of the full details of our approach
until \autoref{sec:Specifying+Formats}. The code accompanying our tour
is included in our code supplement in the
\texttt{src/Narcissus/Examples/README.v} file.

\subsection{A Tour of \frameworkName}
\label{sec:narcissusTour}

\edef\normalindent{\the\parindent}

\paragraph{Getting started}
Our first format is extremely simple:
\NarcissusExampleOneCol{sensor0}

\noindent{}All user input is contained in \autoboxref{sensor0}{1}. \verb|sensor_msg| is a record
type with two fields; the Coq \verb|Record| command defines accessor
functions for these two fields. \verb|format| specifies how instances
of this record are serialized using two format \emph{combinators}:
\verb|format_word| is a \frameworkName primitive that serializes a word
bit-by-bit, and \verb|++| is a sequencing operator (write this, then
that).  \verb|invariant| specifies additional constraints on
well-formed packets, although this example does not have any.  The
\verb|derive_encoder_decoder_pair| tactic is part of the framework and
automatically generates encoder and decoder functions, as well as proofs that they
are correct.

Boxes \boxref{sensor0}{2} and \boxref{sensor0}{3} show the generated code. In \autoboxref{sensor0}{2}, the
encoder operates on a data value and a fixed-size byte buffer (both implicit) and returns the encoded packet, or \verb|None| if it did not fit in the supplied buffer. In \autoboxref{sensor0}{3}, the decoder takes a buffer and returns a packet, or \verb|None| if the buffer did not contain a valid encoding. Both generated programs live in stateful error monads (\verb|<-| and \verb|≫| are the usual binding and  sequencing operators), offering primitives to read and write a single byte (\verb|GetCurrentByte|, \verb|SetCurrentByte|).  The encoder uses the \verb|⋙| reverse-composition operator (\verb|a ⋙ b| $\equiv$ \verb|b ∘ a|) to pass record fields to \verb|SetCurrentByte|.  Since \verb|data| is 16 bits long, the encoder also uses \verb|high_bits| and \verb|low_bits| to extract the first and last 8 bits, and the decoder reassembles them using the \verb|⋅| concatenation operator: this byte-a\-lign\-ment transformation is part of the \verb|derive_encoder_decoder_pair| logic.

\paragraph{Underspecification} We now consider a twist: to align
\verb|data| on a 16-bit boundary, we introduce 8 bits of
padding after \verb|stationID|; these bits will be reserved for future use:
\NarcissusExampleOneCol{sensor1}

\noindent{}These eight underspecified bits introduce an asymmetry: the encoder always writes \verb|0x00|, but the decoder
accepts any value.  The lax behavior is crucial because
the \verb|format_unused_word| specification allows conforming encoders to
output \emph{any} 8-bit value; as a result, a correct decoder for this
format needs to
accept \emph{all} 8-bit values.  In that sense, the encoder and decoder that
\frameworkName generates are not strict inverses of each other: the encoder
is one among many functions permitted by the formatting
specification, and the decoder is the inverse of the \emph{entire
  family} described by the format, accepting packets serialized by
\emph{any} conforming encoder.

\paragraph{Constants and enums} Our next enhancements are to add a version number to our format and to tag each measurement with a \verb|kind|, \verb|"TEMP"| or \verb|"HUMIDITY"|.  To save space, we allocate 2 bits for the tag and 14 bits for the measurement:
\NarcissusExampleOneCol{sensor2}

\noindent{}The use of \verb|format_const| in the specification forces conforming
encoders to write out the value \texttt{0x7e2}, encoded over 16 bits.  Any
input that does not contain that exact sequence is malformed, which
the generated decoder signals by throwing an
exception. \frameworkName also checks more subtle dependencies
between subformats: for example, if a format were to encode the same
value twice, the generated decoder will decode both values \emph{and}
check that they agree--- the packet must be malformed if not.  The argument
passed to \verb|format_enum| specifies which bit patterns to use to
represent each tag (\verb|0b00| for \verb|"TEMP"|, \verb|0b01|
for \verb|"HUMIDITY"|), and the decoder uses this mapping to
reconstruct the appropriate enum member.

\paragraph{Lists and dependencies} Our penultimate example illustrates data dependencies and input
restrictions.  To do so, we replace our single data point with a list
of measurements (for conciseness, we remove tags and use 16-bit
words):
\NarcissusPartialExample{sensor3}

\noindent{}The \verb|format_list| combinator encodes a value by simply
applying its argument combinator in sequence to each element of a
list.  We start a derivation as before, but we quickly run into an
issue: the derivation fails, leaving multiple Coq goals unsolved. The
first of these shows the portion of the format where
\verb|derive_encoder_decoder_pair| got stuck:
\spacedCodeBlock[\centering]{\begin{ESHInlineBlock}[\ESHInlineBlockVerticalAlignment]\input{listings/sensor3fail.v.esh.tex}\unskip\end{ESHInlineBlock}}

\noindent{}Using an additional tactic takes us to the last unsolvable goal, which
is equivalent to the following:
\spacedCodeBlock[\centering]{\begin{ESHInlineBlock}[\ESHInlineBlockVerticalAlignment]\input{listings/sensor3fail2.v.esh.tex}\unskip\end{ESHInlineBlock}}

\noindent{}This goal indicates that the derivation got stuck trying
to find a decoder for the list of measurements. The issue is that the
built-in list decoder is only applicable if the number of elements to
decode is known, but our format never encodes the length of the
\verb|data| list.

An attempt to fix this problem by including the length of \verb|data|
does not completely solve the problem, unfortunately
(\verb|format_nat 8 ◦ length| specifies that the length should
be truncated to 8 bits and written out):

\NarcissusPartialExample{sensor4}

\noindent{}Indeed, the decoder derivation now
gets stuck on the following goal:
\spacedCodeBlock[\centering]{\begin{ESHInlineBlock}[\ESHInlineBlockVerticalAlignment]\input{listings/sensor4fail.v.esh.tex}\unskip\end{ESHInlineBlock}}

\noindent{}Our debugging tactic now produces the following goal:
\spacedCodeBlock[\centering]{\begin{ESHInlineBlock}[\ESHInlineBlockVerticalAlignment]\input{listings/sensor4fail2.v.esh.tex}\unskip\end{ESHInlineBlock}}

\noindent{}The problem is that, since we encode the list's length on 8 bits, the round-trip
property that \frameworkName enforces only holds if the list has fewer
than \(2^{8}\) elements: larger lists have their lengths truncated,
and it becomes impossible for the decoder to know for certain how many
elements it should decode.  What we need is an input restriction: a
predicate defining which messages we may encode.  To this end, we make
one final adjustment:
\NarcissusExampleOneCol{sensor5}
\paragraph{User-defined formats}
Our final example illustrates a key benefit of the combinator-based
approach: integration of user-defined formats and decoders. The
advantage here is that \frameworkName does not sacrifice correctness
for extensibility: every derived function must be correct. This
example uses a custom type for sensor readings, \verb|reading|. To
integrate this type into \frameworkName, the user also supplies the
format specification for this type, corresponding encoders and
decoders and proofs of their correctness, and a set of tactics
explaining how to integrate this record into a
derivation. \autoref{sec:Extensions} provides the complete details on
these ingredients, but for now we note that the \verb|format_reading|
specification is nothing more exotic than a nondeterministic function
in the style of the Fiat framework~\cite{FiatPOPL15}, and that the two
lemmas are normal interactive Coq proofs.
\NarcissusPartialExample{sensor6}

\paragraph{Wrapping up}
In \frameworkName, users specify formats using a library of
combinators, and use tactics to automatically derive
correct-by-construction encoder and decoder functions from these
specifications. Formats may be underspecified, in that a particular
source value may be serialized in different ways, but decoders are
guaranteed to correctly interpret all of them. Formats may induce
dependencies between subformats; the derivation procedure is
responsible for tracking these dependencies when generating a
decoder. Finally, a user can extend \frameworkName with new formats
and datatypes by providing a few simple ingredients; extensions are
guaranteed not to compromise the correctness of derived functions.

To more precisely summarize this paper's contributions:
\begin{itemize}
\item We develop specifications of correctness for encoders and
  decoders, keyed on a common nondeterministic format.
\item We show how encoder and decoder combinators can be verified
  modularly, even when their correctness depends on the contexts in
  which they are used, in a way that enables compositional
  verification of composite encoders and decoders built from
  combinators.
\item We show how to derive correct-by-construction encoders and decoders via
  interactive proof search using libraries of verified combinators,
  and provide proof tactics to automate the process in a way that
  supports extension without compromising soundness.
\item We demonstrate how a two-phase approach which iteratively
  refines bit-level specifications into byte-level functions can
  enable both clean specifications and efficient implementations.
\end{itemize}
\noindent We demonstrate the applicability of \frameworkName by
deriving packet processers for a full Internet protocol stack, which
required the addition of a checksum combinator. Inserting our
replacement into the OCaml-based MirageOS unikernel results in minimal
performance degradation.

We pause briefly here to contrast the design choices made by
\frameworkName with other approaches to serializing and deserializing
data, with a fuller discussion deferred to
\autoref{sec:RelatedWork}. There has been a particular focus on
formally verifying parsers and pretty printers for
programming-language ASTs as parts of compiler
frontends~\cite{Barthwal+2009, Jourdan+2012, Koprowski+2011} or to
carry out binary analysis~\cite{Morrisett+2012, Tan+2018}. One of the
target applications of \frameworkName is formally verified distributed
systems, and the restriction to context-free languages (as found in
those tools) disallows many of the standard network formats such
applications require.

\frameworkName has a similar motivation to bidirectional programming
languages~\cite{Mu+2004, bohannon+2008} in which programs can be run
``in reverse'' to map target values to the source values that produced
them. The bidirectional programming language Boomerang adopts a
similar combinator-based approach to deriving transformations between
target and source values. Invertibility is an intrinsic property of
bidirectional languages, so new combinators require extensions to its
metatheory. In contrast, proofs of correctness are built alongside
functions in \frameworkName, allowing the framework to be safely
augmented by including a proof justifying a new implementation
strategy as part of an extension.

We now present the complete details of \frameworkName in a more
bottom-up fashion, before discussing our evaluation and a more
detailed comparison with related work.  The pieces described below are
contained in our code supplement, which may be helpful to consult
while reading.

\section{Narcissus, Formally}
\label{sec:Specifying+Formats}
We begin our ground-up explanation of \frameworkName with the
definition of the \emph{formats} that capture relationships between
structured source values and their serialized representations. The
signature of a format from source type \verb|S| to target type
\lstinline|T| is defined by a type alias:
\begin{lstlisting}
FormatM S T BIGSIGMA := Set of (S * BIGSIGMA * T * BIGSIGMA)
\end{lstlisting}
\noindent That is, a format is a quaternary relation on source values,
initial states, target values, and final states.  Including states in
the format allows us to specify a rich set of formats, including DNS
packets. As hinted at by the \lstinline|M| suffix, \lstinline|FormatM|
can be interpreted as the composition of the nondeterminism and
stateful monads.

The format combinators showcased in \autoref{sec:narcissusTour} have
straightforward definitions using standard set operations. The
\lstinline|++| combinator sequences its subformats using a monoid
operation \lstinline|BINOP| provided by its target type.
\begin{lstlisting}
(s, SIGMA, t, SIGMA') ELEMENT format_1 ++ format_2 ===
             exists t_1 t_2 SIGMA''. (s, SIGMA, t_1, SIGMA'') ELEMENT format_1 /\ (s, SIGMA'', t_2, SIGMA') ELEMENT format_2 /\ t = t_1 BINOP t_2
\end{lstlisting}
\noindent The function-composition combinator \lstinline|COMPOSE| is
actually defined via the more elementary \lstinline|COMPOSER|
combinator. This combinator uses a relation, \lstinline|f|, to format
a projection of the source domain:
\begin{lstlisting}
(s, SIGMA, t, SIGMA') ELEMENT format COMPOSER f === exists s'. (s', SIGMA, t, SIGMA') ELEMENT format /\ f s s'
\end{lstlisting}
Underspecified formats can be built by combining \lstinline|COMPOSER| with
a choice operator, as in the format for unused words:
\begin{lstlisting}
format_unused_word === format_word COMPOSER {(_, _) | True }
\end{lstlisting}
\noindent In addition to \lstinline|COMPOSE|, \lstinline|COMPOSER| is
used to define the \lstinline|CAP| combinator that restricts the
source values included in a format:
\begin{lstlisting}
format COMPOSE f === format COMPOSER {(s, s') | s' = f s}
format CAP P  === format COMPOSER {(s, s') | P s /\ s = s' }
\end{lstlisting}
Another helpful higher-order combinator is \lstinline|Union|, which is
useful for defining formats with variant encodings, e.g. Ethernet
frames:
\begin{lstlisting}[escapeinside={(*}{*)}]
(s, SIGMA, t, SIGMA') ELEMENT (*$\mathsf{format_1} \cup \mathsf{format_2}$*) === (s, SIGMA, t, SIGMA') ELEMENT (*$\mathsf{format_1}$*) \/ (s, SIGMA, t, SIGMA') ELEMENT (*$\mathsf{format_2}$*)
 \end{lstlisting}
While not very useful for user-defined formats, the empty format
\lstinline|epsilon|, is helpful in the specifications of encoder and decoder combinators:
\begin{lstlisting}
(s, SIGMA, t, SIGMA') ELEMENT epsilon === t = ID /\ SIGMA = SIGMA'
\end{lstlisting}

For clarity, we have presented these combinators in point-free style,
but the monad formed by \lstinline|FormatM| also admits definitions in
a pointed style, which can be more convenient for defining base
formats like \lstinline|format_reading|. In addition to the standard
\lstinline{return} and \lstinline{bind} (\lstinline{ _ <- _; _})
operators, this monad includes a set-comprehension operator
\lstinline!{ x | P x }!, which specifies a set via a defining property
$\mathsf{P}$ on possible return values. The three operators have
straightforward interpretations as sets~\cite{FiatPOPL15}:
\begin{lstlisting}
 e ELEMENT return v === e = v
 e ELEMENT { x | P x } === P e
 e ELEMENT x <- y; k x === exists e'. e' ELEMENT y /\ e ELEMENT k e'
\end{lstlisting}
\noindent As an example, we can specify the set of all possible
locations of a period in a string \lstinline{s} as:
\begin{lstlisting}
  s_1 <- { s_1 : String | exists s_2. s = s_1 ++ "." ++ s_2 }; return (length s_1)
\end{lstlisting}

\begin{figure}
  \centering
  \begin{tabular}{cc}
    \textbf{Format} & Higher-order? \\ \hline
    Booleans & no \\
    Peano Numbers & no \\
    Variable-Length List & yes \\
    Variable-Length String & no \\
    Option Type & yes \\
    Enumerated Types & no
  \end{tabular}\hspace{1em}
  \begin{tabular}{cc}
    \textbf{Format} & Higher-order? \\ \hline
    Fixed-Length Words & no \\
    Unspecified BitString & no \\
    Fixed-Length List & yes \\
    Fixed-Length String & no \\
    Ascii Character & no \\
    Variant Types & yes
  \end{tabular}

  \caption{Formats for base types included in \frameworkName.}
  \label{fig:FrameworkComponents}
  \vspace{-.2cm}
\end{figure}

\begin{wrapfigure}{r}{0.5\linewidth}\vspace{0.5\baselineskip}
  \begin{minipage}{\linewidth}
    \vspace{-.7cm}
\begin{lstlisting}[basicstyle=\footnotesize]
ADT ByteString {
Definition ID : ByteString;
Definition BINOP : ByteString -> ByteString -> ByteString;
Definition snoc : ByteString -> Bool -> ByteString;
Definition unfold : ByteString -> option (Bool * ByteString);
\end{lstlisting}
    \vspace{-.1cm}
\begin{lstlisting}[basicstyle=\footnotesize]
Axiom left_id : forall s_1, ID BINOP s_1 = s_1;
Axiom right_id : forall s_1, s_1 BINOP ID = s_1;
Axiom assoc : forall s_1 s_2 s_3, s_1 BINOP (s_2 BINOP s_3) = (s_1 BINOP s_2) BINOP s_3;
\end{lstlisting}
    \vspace{-.1cm}
\begin{lstlisting}[basicstyle=\footnotesize]
Axiom unfold_app : forall b s_1 s_2 s_3, unfold s_1 = Some (b, s_2)
                  -> unfold (s_1 BINOP s_3) = Some (b, s_2 BINOP s_3);
Axiom snoc_app : forall b s_1 s_2,
                  snoc b (s_1 BINOP s_2) = s_1 BINOP (snoc b s_2);
Axiom unfd_snoc : forall b, unfold (snoc b ID) = Some (b, ID);
Axiom unfd_id : unfold ID = None;
Axiom unfd_inj : forall s_1 s_2, unfold s_1 = unfold s_2 -> s_1 = s_2}
\end{lstlisting}
    \vspace{-.5cm}
  \captionof{figure}{The ByteString interface, with some length operations elided. }
\label{fig:ByteStringADT}
\end{minipage}
\end{wrapfigure}
\noindent In addition to enabling users to define their own formats in
a familiar monadic style, the nondeterminism monad integrates nicely
with Coq's rewriting machinery when deriving correct
encoders. \frameworkName includes a library of formats for the
standard types listed in \autoref{fig:FrameworkComponents}, most of
which have pointed definitions.

We have left the definition of the target type of our formats
underspecified until now. Either \emph{bitstrings} or
\emph{bytestrings}, i.e. lists of bits or bytes, would be natural
choices, each with its own advantages and disadvantages. Bitstrings
have a conceptually cleaner interface which allows users to avoid byte
alignment considerations. As an example, the format in
\autoboxref{sensor2}{1} can simply sequence 14-bit and 2-bit words,
while a byte-aligned specification would require splitting the first
word into 8- and 6-bit words, and combining the latter with the 2-bit
word. Unfortunately, true bitstrings are quite removed from the byte
buffers used in real systems, requiring bit-shifting to enqueue bits
one at a time. \frameworkName attempts to split the difference by
using the bitstring abstract data type presented in
\autoref{fig:ByteStringADT} for the target type of formats. Clients
can treat \lstinline|ByteString| as a bitstring equipped with
operations governed by algebraic laws for monoids and queues, while
its actual representation type is closer to that of a
bytestring. \autoref{sec:synthesis} details how our derivation
procedures optimize away uses of this interface in order to produce
more performant implementations.

\subsection{Specifying Encoders and Decoders}
\label{sec:VerifiedEncoders+Decoders}
These relational formats are not particularly useful by themselves---
even checking whether a format permits specific source and target
values may be undecidable. Instead we use them to specify the
correctness of \emph{both} encoders and decoders.
So far, we have seen examples of relational formats that permit one or
many target representations of a particular source value, but in their
full generality, there might not be \emph{any} valid encodings of some
source value. As an extreme example, consider the following use of the
\lstinline|CAP| combinator to define an empty relation:
\lstinline!{s | False} CAP format_word!.
More realistically, a format for domain names must disallow strings with
runs of ``.'', e.g. ``www.foo..bar.com''. To account for formats
that exclude some source values, \frameworkName encoders are
partial functions from source to target values:
\lstinline|EncodeM S T BIGSIGMA := S -> BIGSIGMA -> Option (T * BIGSIGMA)|.
At a high level, a format describes a family of permissible encoders, where
each must commit to a single target representation for each source value in
the relation. More formally:
\begin{definition}[Encoder Correctness]
  A correct encoder for a format
  \lstinline|format : FormatM S T BIGSIGMA| is a partial function,
  \lstinline|encode : EncodeM S T BIGSIGMA|, that only produces
  encodings of source values included in the format and produces
  an error on source values not included in the format:
  \begin{align*}
    & \forall s\, \sigma\, t\, \sigma'. \;
      \mathsf{encode}\; s\; \sigma = \mathsf{Some}\;(t,
    \sigma') \rightarrow (s, \sigma, t, \sigma') \in \mathsf{format} \\
    \bigwedge & \forall s\, \sigma. \; \mathsf{encode}\;s\; \sigma = \bot
                \rightarrow \forall t'\; \sigma'.\; (s, \sigma, t',
                \sigma') \not \in \mathsf{format}.
  \end{align*}
\end{definition}
\noindent In other words, a valid encoder \emph{refines} a format; we
henceforth use the notation \lstinline|format REFINEDBY encode| to
denote that \lstinline|encode| is a correct encoder for
\lstinline|format|.

Before stating the corresponding correctness definitions for decoders,
consider the high-level properties a correct decoder should satisfy,
ignoring for now the question of state. Clearly, it must be a
\emph{sound} left inverse of the format relation. That is, it should
map every element $t$ in the image of $s$ in \lstinline|format| back
to $s$:
$\forall s \; t. \; (s, t) \in \mathsf{format} \rightarrow
\mathsf{decode}(t) = s$. Less clear is how much conformance
checking a ``correct'' decoder should perform on target values
that fall outside the image of \lstinline|format|: should
\lstinline|decode| fail on such inputs, or should its behavior be
unconstrained in these cases? If these decoders are being integrated
into other formally verified systems that process decoded data
further, it is desirable to provide the strongest assurance about the
integrity of decoded data to downstream functions. On the other hand,
there are valid reasons for looser standards, like sacrificing
strict format validation for efficiency,
e.g. by not verifying checksums. For now, we require correct decoders
to flag strictly \emph{all} malformed target
values by signaling errors when applied to target values not
included in the relation:
$\forall s \, t, \; \mathsf{decode}\; t = \mathsf{Some}\; s
\rightarrow (s, t) \in \mathsf{format}$. As we shall see later, our
formulation will also support deriving decoders with more lenient
validation policies. However, we note that there are compelling
security reasons for the top-level decoder to enforce strict input
validation, in order to cut off potential side channels via demonic
choice between legal alternatives.  (E.g., consider a decoder
integrated within an e-mail server, which decodes malformed packets
into the contents of other users' inboxes.)

Looking at the signature of decoders in \frameworkName,
\begin{lstlisting}
DecodeM S T BIGSIGMA := T -> BIGSIGMA -> Option (S * BIGSIGMA)
\end{lstlisting}
we see that we need to adapt these notions of correctness to account for
the state used by a decoder.  Whereas an encoder is a refinement of a
format, and thus used identical types of state, we do not force
compatible decoders and formats to share the same type of state. To
see why, consider a simplified version of the format for DNS domain
names~\cite{RFC1035}, which keeps track of the locations of previously
formatted domains via its state argument:
\begin{lstlisting}[escapeinside={(*}{*)}]
Let format_domain (d: domain) (SIGMA: domain -> word) := format_name d UNION format_word (SIGMA d)
\end{lstlisting}
\noindent This example uses an optional compression strategy in which
a domain name can either be serialized or replaced with a pointer to
the location of a previously formatted occurrence. A decoder for
domains should also keep track of this information, in order to decode
pointers:
\begin{lstlisting}
Let decode_domain (t : T) (SIGMA : word -> domain) := ...
\end{lstlisting}
In order for \lstinline|decode_domain| to be correct, its state needs
to ``agree'' with the state used to format its input. We do not want to
require that these states be equal, so that decoders can have the freedom
to use different data structures than the format,
e.g. \lstinline|decode_domain| could be implemented using a BST sorted
on words, while \lstinline|format_domain| could use a prefix trie on
domain names. Our notion of decoder correctness captures agreement
between different state types via a binary relation which defines
when format and decoder states are consistent. Hence our full notion
of decoder correctness accounts for both state and erroneous
target values.

\begin{definition}[Decoder Correctness]
  A correct decoder for a format,
  \lstinline|format : FormatM S T BIGSIGMAE|, and relation on states,
  \lstinline|EQUIV : Set of (BIGSIGMAE * BIGSIGMAD)|, is a function,
  \lstinline|decode : DecodeM S T BIGSIGMAD|, that, when applied to a
  valid target value and initial state, produces a source value and
  final state similar to one included in \lstinline|format|,
  signaling an error otherwise: \\
\begin{tabular}{c}
$\Bigg(~$
\begin{lstlisting}
forall(SIGMA_E SIGMA_E': BIGSIGMAE) (SIGMA_D:BIGSIGMAD) (s : S) (t : T).
  (s, SIGMA_E, t, SIGMA_E') ELEMENT format /\ SIGMA_E EQUIV SIGMA_D
-> exists SIGMA_D'. decode t SIGMA_D = Some (s, SIGMA_D')
       /\ SIGMA_E' EQUIV SIGMA_D'
\end{lstlisting} $~\Bigg)$
$\bigwedge$
$\Bigg(~$
\begin{lstlisting}
forall(SIGMA_E:BIGSIGMAE) (SIGMA_D SIGMA_D':BIGSIGMAD) (s : S) (t: T).
  SIGMA_E EQUIV SIGMA_D /\ decode t SIGMA_D = Some (s, SIGMA_D')
-> exists SIGMA_E'. (s, SIGMA_E, t, SIGMA_E') ELEMENT format /\ SIGMA_E' EQUIV SIGMA_D'
\end{lstlisting}
$~\Bigg)$
\end{tabular}
\end{definition}
\noindent We denote that \lstinline|decode| is a correct
decoder for \lstinline|format| under a similarity relation on states
\lstinline|EQUIV| as \lstinline|format| $\RoundTripMathAp$
\lstinline|decode|.

By definition, it is impossible to find a correct decoder for a
non-injective format. While decoders and encoders have independent
specifications of correctness, using a common format provides a
logical glue that connects the two. We can, in fact, prove the
expected round-trip properties between a correct encoder and correct
decoder for a common format:

\begin{theorem}[Decode Inverts Encode]
  Given a correct decoder
  $\mathsf{format} \RoundTripMathAp \mathsf{decode}$ and correct
  encoder $\mathsf{format} \supseteq \mathsf{encode}$ for a common
  format \textup{\lstinline|format|}, \textup{\lstinline|decode|} is an inverse for
  \textup{\lstinline|encode|} when restricted to source values in the format:
  \[\forall s\, \sigma_E \, t\, \sigma_E' \, \sigma_D. \;
    \mathsf{encode}\;s\;\sigma_E = \mathsf{Some}(t, \sigma_E')~\land~
    \sigma_E \approx \sigma_D \rightarrow \exists \sigma_D'. \;
    \mathsf{decode}\; t\;\sigma_D = \mathsf{Some}(s, \sigma_D')\]
\end{theorem}

\begin{theorem}[Encode Inverts Decode]
  \label{thm:EncodeInvertsDecode}
  Given a correct decoder
  $\mathsf{format} \RoundTripMathAp \mathsf{decode}$ and correct
  encoder $\mathsf{format} \supseteq \mathsf{encode}$ for a common
  format \textup{\lstinline|format|}, \textup{\lstinline|encode|} is defined for all
  decoded source values produced by \textup{\lstinline|decode|},
  \[\forall s\, \sigma_D \, t\, \sigma_D' \, \sigma_E. \;
  \mathsf{decode}\; t\;\sigma_D = \mathsf{Some}(s, \sigma_D') \land
  \sigma_E \approx \sigma_D \rightarrow
  \exists t'\, \sigma_E'. \; \mathsf{encode}\;s\;\sigma_E = \mathsf{Some}(t', \sigma_E')\]
\end{theorem}
That \lstinline|encode| is an inverse of \lstinline|decode| for source
values with unique encodings is a direct corollary of
\autoref{thm:EncodeInvertsDecode}.

\section{Deriving Encoders and Decoders}
\label{sec:CorrectEncoders}
Equipped with precise notions of correctness, we can now define how we
derive provably correct encoders and decoders from a format. These
functions will be byte-aligned in a subsequent derivation step
presented in \autoref{sec:ByteAlignment}. We begin with encoders, since
they often have similar structure to their corresponding
formats. Intuitively, such a derivation is simply the search for a
pair of an encoder function \lstinline|encode| and a proof term
witnessing that it is correct with respect to a format:
\lstinline|format REFINEDBY encode|. As an example, a proof that a
function which returns an empty bytestring correctly implements the
empty format can also be read as evidence that it is safe to choose
this implementation when searching for an encoder for
\lstinline|epsilon|. In this light, lemmas like
\lstinline|enc_readingCorrect| which prove that encoder combinators
are correct can be interpreted as \emph{derivation rules} for
constructing such proof trees from goal formats.

Leveraging this intuition, we denote these lemmas using standard
inference-rule notation:
\begin{center}
\begin{tabular}{rcl}
\begin{lstlisting}[escapeinside={(*}{*)}]
Lemma EncA (h(*$_1$*) : H_1) (h(*$_2$*) : H_2)
 : CorrectEncoder A T BIGSIGMA format_A  encode_A.
\end{lstlisting}
& \scalebox{1.5}{$\equiv$} &
\begin{minipage}{.27\linewidth}
\infrule[EncA]
{
  \hbox{\lstinlines|H_1|}
  \andalso \hbox{\lstinlines|H_2|}
}
{
  \hbox{\lstinlines|format_A|}
  \supseteq
  \hbox{\lstinlines|encode_A|}
}
\end{minipage}
\end{tabular}

\end{center}

\begin{figure}[t]

  \infrule[EncSeq]
{
  \hbox{\lstinlines|format_1 REFINEDBY encode_1|}
  \andalso \hbox{\lstinlines|format_2 REFINEDBY encode_2|} \\
    \lstinlines|forall (s, SIGMA, t_1, SIGMA') ELEMENT format_1.|
    \lstinlines|forall t' SIGMA''. (s, SIGMA', t', SIGMA'') ELEMENT format_2 ->|\\
    \phantom{\lstinlines|forall (s, SIGMA, t_1, SIGMA') ELEMENT
      format_1.|}
    \lstinlines|exists t_1' SIGMA_2 t_2' SIGMA_3. encode_1 s SIGMA = Some (t_1', SIGMA_2)|
    \lstinlines| WEDGE encode_2 s SIGMA_2 = Some (t_2', SIGMA_3) |
}
{
  \hbox{\lstinlines|format_1 ++ format_2|}
  \supseteq
  \hbox{\lstinlines|FUN's.| \lstinlines|t_1 <- encode_1 s;|
    \lstinlines|t_2 <- encode_2 s;|
    \lstinlines|return (t_1 BINOP t_2)|}
  }
\begin{tabular}{cc}
  \begin{minipage}{.35 \linewidth}
\infrule[EncComp]
{
  \hbox{\lstinlines|format|}
  \supseteq
  \hbox{\lstinlines|encode |}
}
  {
    \hbox{\lstinlines|format COMPOSE g|}
    \supseteq
    \hbox{\lstinlines|encode COMPOSE g |}
}
\end{minipage}

  & \begin{minipage}{.6 \linewidth}
    \infrule[EncRest]
    {
    \hbox{\lstinlines|format REFINEDBY encode|}
    \andalso
    \hbox{\lstinlines|forall s. p s = true <-> s ELEMENT P |}
    }
    {
      \hbox{\lstinlines|P CAP format|}
      \supseteq
      \hbox{\lstinlines|FUN's. if p s then encode s else fail|}
}
\end{minipage}
  \\

  \begin{minipage}{.35 \linewidth}
    \vspace{.2cm}
  \infrule[EncEmpty]
{
}
{
  \hbox{\lstinlines|epsilon|}
  \supseteq
  \hbox{\lstinlines|FUN's. return ID |}
}
\end{minipage}
  &
                 \begin{minipage}{.6 \linewidth}
                   \vspace{.2cm}
\infrule[EncUnion]
{ \hbox{\lstinlines|format|}_\mathsf{1}
  \supseteq \hbox{\lstinlines|encode|}_\mathsf{1}
  \andalso
  \hbox{\lstinlines|format|}_\mathsf{2}
  \supseteq
  \hbox{\lstinlines|encode|}_\mathsf{2}
  \\
  \hbox{\lstinlines|forall (s, SIGMA, t, SIGMA')|} \in
  \lstinlines|format|_\mathsf{i}. \;
  \lstinlines|n s| = \lstinlines|i|
}
{
  \lstinlines|format|_\mathsf{1} \cup \lstinlines|format|_\mathsf{2}
  \supseteq
  \hbox{\lstinlines|FUN's. j <- n s; encode|}_\mathsf{j} \hbox{\lstinlines|s|}
}
\end{minipage}
\end{tabular}

\caption{Correctness rules for encoder combinators.}
\label{fig:CorrectFormatLemmas}
\end{figure}

\autoref{fig:CorrectFormatLemmas} presents the encoder combinators for
the formats from \autoref{sec:Specifying+Formats} using this
inference-rule style. \textsc{EncUnion} is an example of an encoder
that commits to a particular target value-- given a correct encoder
for each format in the union, it relies on an index function
\lstinline|n| on source values to commit to a particular encoding
strategy, with the second hypothesis of the rule ensuring that this
index function correctly picks a format that includes the source
value. The rule for \lstinline|++| proves that a correct encoder for
sequences can be built from encoders for its subformats.  This rule
features a wrinkle concerning state: in order to apply \textsc{EncSeq}
correctly, if the \lstinline|format_1| produces \emph{some}
intermediate state that makes \lstinline|format_2| (and thus
\lstinline|format_1 ++ format_2|) nonempty, \lstinline|encode_1| must
also produce a state on which \lstinline|encode_2| returns
some value. The last hypothesis of \textsc{EncSeq}
enforces that \lstinline|encode_1| does not ``mislead''
\lstinline|encode_2| in this manner.

\begin{figure}[h]
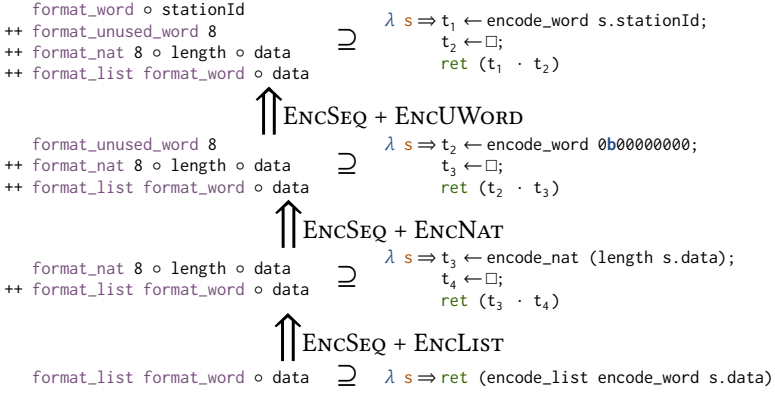

  \def\ESHInlineBlockFontSize{\scriptsize}
  \newcommand{\refinementFig}[1]{%
    \begin{ESHInlineBlock}[\ESHInlineBlockVerticalAlignment]\input{listings/#1spec.v.esh.tex}\unskip\end{ESHInlineBlock} &
    \scalebox{1.25}{$\supseteq$} &
    \begin{ESHInlineBlock}[\ESHInlineBlockVerticalAlignment]\input{listings/#1impl.v.esh.tex}\unskip\end{ESHInlineBlock}\\
  }
  \newcommand{\refinementArrow}[1]{%
    \multicolumn{3}{c}{\scalebox{2}{$\Uparrow$}\textsc{#1}}\\
  }
  \centering
  \begin{tabular}{@{}rcl@{}}
    \centering
    \refinementFig{derivation0}
    \refinementArrow{EncSeq + EncUWord}
    \refinementFig{derivation1}
    \refinementArrow{EncSeq + EncNat}
    \refinementFig{derivation2}
    \refinementArrow{EncSeq + EncList}
    \refinementFig{derivation3}
  \end{tabular}
  \caption{An example encoder derivation.}
  \label{fig:EncoderExample}
  \vspace{-1\baselineskip}
\end{figure}

By combining \textsc{EncSeq} and \textsc{EncComp} with the rules for
base types, e.g. \textsc{EncWord}, we can iteratively derive correct
encoders from a format; \autoref{fig:EncoderExample} presents an
example of such a derivation for one of the encoders from our
introductory tour. Each step in the derivation corresponds to the
encoder that results from applying \textsc{EncSeq} and an
encoder-derivation rule for the topmost format. The proofs that none
of the encoders pass on misleading states are elided. The hole
$\square$ at each step corresponds to the encoder that is built at the
next step; recursively filling these in and simplifying the resulting
expression with the monad laws yields the expected encoder for
\lstinline|enc_data|.

\subsection{Decoders}
\label{sec:CorrectDecoders}
Before defining similar correctness rules for decoder
combinators, we pause to consider how they are used to build a
top-level decoder. In particular, consider what the decoder
combinators used to build a reusable decoder for \lstinline|++| should
look like:
\[\mathsf{format_1 +\hspace{-3pt}+\, format_2}~ \RoundTripMathAp
  \mathsf{decode_1 >\!\!>\!\!= decode_2} \]
The natural way to decode the value resulting from sequencing
$\mathsf{format_1}$ and $\mathsf{format_2}$ is to have
$\mathsf{decode_1}$ return any unconsumed portion of the target value
for $\mathsf{decode_2}$ to finish processing. We thus define the signature of
a decoder combinator to be:
\begin{lstlisting}
DecodeCM V T BIGSIGMA := T -> BIGSIGMA -> Option (V * T * BIGSIGMA)
\end{lstlisting}
The change of the name of the first type parameter also suggests a
more subtle difference between intermediate and top-level decoders, in
that the former return partial projections or \emph{views} of
source values used to produce the target inputs. Unfortunately, neither
of these changes align with our earlier notion of decoder
correctness, which expects a decoder to recover the \emph{full} source
value by completely consuming a target value. To recover the desired
top-level property, we first adapt our two correctness
properties to account for these differences. To concretize this
discussion, consider how we might justify the use of
\lstinline|decode_word| in the following decoder correctness fact.
\begin{center}
\begin{tabular}{rcl}
\begin{lstlisting}[escapeinside={(*}{*)}]
   format_word COMPOSE stationId
++ format_word COMPOSE data
\end{lstlisting} &
                   $\roundtripAp $
  &
\begin{lstlisting}[escapeinside={(*}{*)}]
id <- decode_word;
d <- decode_word;
return { stationId := sid; data := d }
\end{lstlisting}
\end{tabular}
\end{center}
The first and second uses of \lstinline|decode_word| compute
projections of the original sensor value, \lstinline|stationId| and
\lstinline|data|, respectively. In addition, note that each combinator
can only validate that its target value is consistent with its view
of the data, not the full source value.  Finally, in order to be
sequenced correctly, a combinator needs to consume \emph{precisely} the
portion of the bitstring corresponding to its projection.  In order
to account for the first two concerns, our adaptation of the soundness
(left-inverse) criterion for decoder combinators is parameterized over
a binary relation on source and projected values representing a
\emph{view}, as well as an additional format capturing the conformance
checking performed by the decoder.

\begin{definition}[Decoder Combinator Soundness]
  A sound decoder combinator for a source format,
  \lstinline|format_s : FormatM S T BIGSIGMAE|, relation on states,
  \lstinline|EQUIV : Set of (BIGSIGMAE * BIGSIGMAD)|, view
  \lstinline|view : S -> V -> Prop|, and conformance format,
  \lstinline|format_v : FormatM V T BIGSIGMAE| is a function,
  \mbox{\lstinline|decode : DecodeCM V T BIGSIGMAD|,} that when applied to a
  valid target value appended to an arbitrary bitstring and initial
  state, produces a view of the source value that agrees with the
  conformance format while consuming exactly the
  portion of the target value in the source format:
\begin{lstlisting}
forall(SIGMA_E SIGMA_E': BIGSIGMAE) (SIGMA_D:BIGSIGMAD) (s : S) (t t': T). (s, SIGMA_E, t, SIGMA_E') ELEMENT format_s /\ SIGMA_E EQUIV SIGMA_D
-> exists v SIGMA_D'. decode (t BINOP t') SIGMA_D = Some (v, t', SIGMA_D') /\
           view s v /\ (v, SIGMA_D, t, SIGMA_D') ELEMENT format_v /\ SIGMA_E' EQUIV SIGMA_D'
\end{lstlisting}
\end{definition}

To see why combinators are required to be oblivious to tails
of bitstrings, consider a simple format for card suits which uses unit
for the state type (we elide the trivial state values below):
\[\mathsf{format\_suit \equiv \{ (\clubsuit, 0b11), (\diamondsuit,
    0b0), (\heartsuit, 0b1), (\spadesuit, 0b10) \}}\]
\noindent To format a pair of cards, we could format each card in
sequence:
\lstinline|format_suit COMPOSE fst ++ format_suit COMPOSE snd|.  This
format is clearly not injective, as it is not possible to distinguish
between the encodings of $\mathsf{(\clubsuit, \diamondsuit)}$ and
$\mathsf{(\heartsuit, \spadesuit)}$. Absent additional information in
the surrounding format, e.g. a Boolean flag identifying the color of
the suit being decoded, it is impossible for a combinator for
\lstinline|format_suit| to identify soundly how much of the target to
process. Since this format lacks such information, it is imposible to
find a correct \emph{top-level} decoder for it.

Adapting the conformance-checking criterion for decoder combinators is
more straightforward. Absent a complete view of the original source
value, a combinator will be unable to ensure adherence to the original
format, but it can ensure that any computed value agrees with the
provided view format relation \emph{and} is a consistent
view of any source values in the original format with the same encoding:
\begin{definition}[Decoder Combinator Consistency]
  A consistent decoder combinator for a source format,
  \lstinline|format_s : FormatM S T BIGSIGMAE|, conformance format,
  \lstinline|format_v : FormatM V T BIGSIGMAE|, relation on states,
  \lstinline|EQUIV : Set of (BIGSIGMAE * BIGSIGMAD)|, view
  \lstinline|view : S -> V -> Prop|, is a function,
  \lstinline|decode : DecodeCM V T BIGSIGMAD|, that is guaranteed to
  produce a view and unconsumed bitstring in a manner consistent
  with the conformance format:
\begin{lstlisting}
forall(SIGMA_E:BIGSIGMAE) (SIGMA_D SIGMA_D':BIGSIGMAD) (v : V) (t t': T). SIGMA_E EQUIV SIGMA_D /\ decode t SIGMA_D = Some (v, t', SIGMA_D')
-> exists t'' SIGMA_E'. t = t'' BINOP t'  /\ (v, SIGMA_E, t'', SIGMA_E') ELEMENT format_v /\ SIGMA_E' EQUIV SIGMA_D'
           /\ forall s. (s, SIGMA_E, t'', SIGMA_E') ELEMENT format_s -> view s v
\end{lstlisting}
\end{definition}

\noindent A correct decoder combinator is one that is both sound with
respect to the source format and a view \lstinline|v| and that is
consistent with its conformance format. We denote this property as
\lstinline|format_s| \roundtripPart{v} \lstinline|decode ~~ format_v|.
Note that choosing the equality relation as the view and the original
format as the conformance format yields a property equivalent to our original
left-inverse criterion, which we continue to denote as
\roundtripAp. Similarly, choosing the relation $\{(s, v) \mid
\mathsf{True} \}$ as the conformance format permits the decoder to return
\emph{any} value when applied to a malformed input.

\autoref{fig:DerivedCorrectDecoderLemmas} presents a selection of some
(strict) decoder combinator correctness theorems included in
\frameworkName as inference rules. Note that all three rules are
parameterized over a predicate \lstinline|P| which restricts the
source format \lstinline|P CAP format|. This predicate is key to our
approach to the modular verification of decoder combinators: each of
these proofs use this predicate to thread information about previously
decoded data through a proof of correctness for a composite
decoder. Such information is necessary for a decoder combinator whose
correctness \emph{depends} on context in which it is used. As a
concrete example, the \lstinline|decode_list| combinator in
\textsc{DecList} only correctly decodes lists of length \lstinline|n|,
a restriction enforced by the second assumption of the rule. In
isolation this rule only justifies using \lstinline|decode_list| to
decode fixed-lengths lists. When used as part of a larger format that
also includes the length of the original list, however, it can be
applied to lists of variable length, as in
\autoboxref{sensor6}{1}. At first glance, the \textsc{DecDone} rule
seems even more limited, as it only applies to a format with a single,
unique source value. In the context of a larger format, however, this
rule becomes much more powerful, particularly when employing these
rules to \emph{derive} a decode from a format specification.

\begin{figure}[!t]
\begin{tabular}{cc}
  \begin{minipage}{.55\textwidth}

    \infrule[DecList]
    {
      \lstinlines|Q CAP format| \RoundTripMathAp \lstinlines|decode| \\
      \mathsf{\scriptstyle \forall l.\; P\; l \rightarrow \mid l \mid =
        n \land \forall a\in\; l. Q\; a}
    }
    {
      \lstinlines|P CAP format_list format|~
      \RoundTripMathAp
      \lstinlines|decode_list decode n|
    }
     \end{minipage}
  &
      \begin{minipage}{.4\textwidth}
  \infrule[DecDone]
{
\mathsf{\scriptstyle \forall s'. \; P\;s'~\rightarrow~ s\;=\;s'} \\
\mathsf{\scriptstyle b = true\; \leftrightarrow \; P\;s}
}
{
  \lstinlines|P CAP epsilon|~
  \RoundTripMathAp
  \lstinlines|if b then return s else fail|
}

\end{minipage}
\end{tabular}

\infrule[DecSeqProj]
{
  \hbox{\lstinlines|Q CAP format_1| }
  \RoundTripMathAp
  \lstinlines|decode_1|
  \andalso
  \mathsf{\scriptstyle \forall s.\; P\;s~ \rightarrow~ Q\; (f\; s)}
  \\
  \mathsf{\scriptstyle \forall v. \;
    \{s~\mid~P\;s\; \land\; f\; s ~=~ v \}~\cap~\lstinlines|format_2| }
  \RoundTripMathAp
  \lstinlines|decode_2 v|
}
{
  \hbox{\lstinlines|P CAP format_1 COMPOSE f ++ format_2|}
  \RoundTripMathAp
  \lstinlines|decode_1 >>= decode_2|
}

\vspace{-.2cm}
\caption{Selected correctness rules for decoder combinators.}
\label{fig:DerivedCorrectDecoderLemmas}
\vspace{-.35cm}
\end{figure}

To see how, consider how the source predicate evolves during the
decoder derivation presented in \autoref{fig:SimpleDerivation}. Each
intermediate node in this derivation corresponds to the format in the
last premise of \textsc{DecSeqProj}. Note how each step introduces a
variable for the newly parsed data, and how an additional constraint
is added to \lstinline|P| relating the original source value to this
value. When the derivation reaches \lstinline|format_list|, this
constraint witnesses that the number of elements in that list is
known. Similarly, although the format is empty at the topmost leaf of
the derivation tree, \lstinline|P| includes enough constraints to
uniquely recover the original source value, and \textsc{DecDone} can
be applied to finish the derivation. The first premise of
\textsc{DecDone} ensures that the restriction on source values is
sufficient to prove the existence of some constant \lstinline|s| that
is equal to the original source value. The second premise of
\textsc{DecDone} ensures that a derived decoder is not overly
permissive in the case that \lstinline|P| is too restrictive. As
previously noted, a format could encode the same view of a source
value twice, and the consistency of the corresponding decoded values
should be validated during decoding. Thus, the function \lstinline|b|
in this premise acts as a decision procedure that validates the
consistency of all the projections of the original source gathered
during decoding. While this example is straightforward, similar
dependencies can be found in many existing binary formats, in the form
of tags for sum types, version numbers, and checksum fields. In each
case, the correctness of a combinator for a particular subformat
depends on a previously decoded value.

\begin{figure}[t]
  \def\ESHInlineBlockFontSize{\small}
  \newcommand{\constraintInter}[1]{%
    \begin{ESHInlineBlock}[\ESHInlineBlockVerticalAlignment]\input{listings/#1cond.v.esh.tex}\unskip\end{ESHInlineBlock} &
    $\cap$ &
    $\left(\text{\begin{ESHInlineBlock}[\ESHInlineBlockVerticalAlignment]\input{listings/#1fmt.v.esh.tex}\unskip\end{ESHInlineBlock}}\right)$\\
  }
  \newcommand{\constraintArrow}[1]{
    \multicolumn{3}{c}{\scalebox{2}{$\Uparrow$}\textsc{#1}}\\
  }
\centering
\begin{tabular}{@{}rcl@{}}
  \constraintInter{constraints0}
  \constraintArrow{DecSeqProj + DecWord}
  \constraintInter{constraints1}
  \constraintArrow{DecSeqUn + DecWord}
  \constraintInter{constraints2}
  \constraintArrow{DecSeqProj + DecNat}
  \constraintInter{constraints3}
  \constraintArrow{DecSeqProj + DecList}
  \constraintInter{constraints4}
\end{tabular}
\caption{An example of constraints added to a format during a
  decoder derivation.}
\label{fig:SimpleDerivation}
\end{figure}

\begin{figure}[!t]
  \vspace{-.1cm}

  \begin{tabular}{cc}
    \begin{minipage}{.45\textwidth}
    \infrule[DecCompose]
{
  \lstinlines|Q CAP format_s| ~
  \RoundTripMathAp
  \lstinlines|decode|
  \\ \mathsf{\scriptstyle \forall s\; v. \; \rho \;s\; v}
  \mathsf{\scriptstyle~~\land~~ P\;s~\rightarrow~Q \;v}
}
{
  \lstinlines|P CAP format_s COMPOSER RHO|~
  \roundtripPartMath{$\rho$}
  \lstinlines|decode ~~ format_v CAP Q|
    }
    \end{minipage}
&
\begin{minipage}{.45\textwidth}

  \infrule[DecViewDone]
{
\mathsf{\scriptstyle \forall s. \; P~s~\rightarrow~\rho\; s\; v} \\
\mathsf{\scriptstyle \forall \sigma. \; (v,\; \sigma,\; \iota,\; \sigma) \in}
\mathsf{\scriptstyle format_v}\
}
{
  \lstinlines|P CAP epsilon|~
  \roundtripPartMath{$\rho$}
  \lstinlines|return v ~~ format_v|
}
\end{minipage}
  \end{tabular}
    \infrule[DecInject]
{
  \lstinlines|P CAP format_s| ~
  \roundtripPartMath{$\rho$}
  \lstinlines|decode ~~ format_v|
  \\ \mathsf{\scriptstyle \forall s\; v. \; \rho \;s\; v }
  \mathsf{\scriptstyle~~\land~~ P\;s~\rightarrow~ \rho\; s\; (f\;v)}
}
{
  \lstinlines|P CAP format_s|~
  \roundtripPartMath{$\rho$}
  \lstinlines|fmap f decode ~~ format_v COMPOSER f|
    }

  \infrule[DecSeq]
{
  \hbox{\lstinlines|P CAP format_1| }
  \roundtripPartMath{$\rho_1$}
  \lstinlines|decode_1 ~~ format_v1|
  \andalso
  \mathsf{\scriptstyle \forall s\; v_1 \; v_2.~ \rho_3\; s\; (v_1, v_2) ~~\leftrightarrow~~
    \rho_1\; s\; v_1 ~\land~ \rho_2\; s\; v_2}
  \\
  \mathsf{\scriptstyle \forall v. \;
    \{s~\mid~P\;s\; \land\; \rho\;s\;v \}~\cap~\lstinlines|format_2| }
  \roundtripPartMath{$\rho_2$}
  \lstinlines|decode_2 v ~~ format_v2 v|
}
{
  \hbox{\lstinlines|P CAP format_1 ++ format_2|}
  \roundtripPartMath{$\rho_3$}
  \lstinlines|v_1 <- decode_1; v_2 <- decode_2; return (v_1, v_2) ~~ format_v1 ++ format_v2|
}

\infrule[DecUnion]
{
  \mathsf{\scriptstyle P\,\cap\, format_T
    \roundtripPartMath{$\rho$}
    decode_T ~\sim~ format_{vT}}
  \andalso
  \mathsf{\scriptstyle P\,\cap\, format_E
    \roundtripPartMath{$\rho$}
    decode_E ~\sim~ format_{vE}}
  \\
  \mathsf{\scriptstyle P\,\cap\, subformat
    \roundtripPartMath{$\rho_B$}
    decode_B ~\sim~}
  \Bigg\{\begin{minipage}{.34\textwidth}
    $\mathsf{\scriptstyle
      (b,\; \sigma,\; t,\; \sigma') \mid
      \forall s\; t'\; \sigma''.\;
    }$
    $\mathsf{\scriptstyle (s,\; \sigma,\; t ++t',\;
      \sigma'') ~\in~ format_T ~\rightarrow~ b = true}$
    $\mathsf{\scriptstyle \land~
      (s,\; \sigma,\; t ++t',\; \sigma'') ~\in~ format_E
      ~\rightarrow~ b = false }$
  \end{minipage}\Bigg\}
  \andalso
  \mathsf{\scriptstyle subformat\; \le \; (format_1 ~\cup~ format_2 )}
}
{
  \mathsf{\scriptstyle P~ \cap~ (format_T ~\cup~ format_E)~}
  \roundtripPartMath{$\rho$}
  \hbox{\lstinlines|FUN't. b <- decode_B; if b then decode_T else
    decode_E ~~|} \mathsf{\scriptstyle ~~ format_{vT} ~\cup~ format_{vE}}
}

\vspace{-.8cm}
\caption{Additional decoder combinator correctness rules.}
\label{fig:CorrectDecoderLemmas}
\vspace{-.35cm}
\end{figure}

\textsc{DecCompose} and \textsc{DecDone} can actually be derived from
the more general set of rules found in
\autoref{fig:CorrectDecoderLemmas}. While these rules are mostly
helpful for proving more specific rules which are more useful in
derivations, each demonstrates some interesting feature of our
formulation of correctness for decoder combinators:
\begin{itemize}
\item \textsc{DecCompose} proves how to correctly decode projections of
  the source value. This rule requires the use of the more general
  correctness statement, as \lstinline|decode| can only recover the
  view of the source value it has access to.
\item \textsc{DecInj} proves how to safely transform a projected
  value, and must update the conformance format to reflect that a
  transformation has been applied. While not particularly helpful
  during derivations, this rule is useful for proving other derivation
  rules.
\item \textsc{DecViewDone} generalizes \textsc{DecDone} to arbitrary
  views of a source value. The second premise corresponds to the
  decision procedure from \textsc{DecDone} --- an empty conformance
  format is one consequence of the source value projections gathered
  during decoding being inconsistent.
\item \textsc{DecSeq} is a mostly straightforward generalization of
  \textsc{DecSeqProj}, with the important tweak that it builds a
  decoder that constructs a pair of the views produced by its
  subdecoders. Composing this rule with \textsc{DecInj} justifies the
  correctness of combinators that drop intermediate views, e.g.
  \textsc{DecSeqProj}.
\item \textsc{DecUnion} is similar to \textsc{UnionEnc}, with the key
  difference being that it requires a boolean value, \lstinline|b|,
  indicating which format produced the current bitstring. The
  combinator uses a decoder, $\mathsf{decode_B}$, to compute this
  value, and uses a conformance check on the result of that decoder to
  ensure the boolean flag is correct. In addition, the proof of
  correctness for $\mathsf{decode_B}$ only requires that it consume
  some prefix of the current format, giving it the freedom to return
  as soon as it can identify which subformat was used to generate the
  current source value. Framing the problem in this way allows
  \frameworkName to leverage other derivation rules to build this
  function.  We will see another example of this paradigm in the rule
  for IP checksums presented in \autoref{sec:Extensions}.
\end{itemize}

\subsection{Improving Performance of Encoders and Decoders}
\label{sec:ByteAlignment}
The encoders and decoders derived via our combinator rules utilize the
same bitstring abstract data type as format specifications, employing
the bitstring's \lstinline|snoc| and \lstinline|unfold| operations
to enqueue and dequeue individual bits. Operating at the bit-level
imposes a large performance hit on these functions, since implementing
these methods on the fixed-length byte buffers typically used for the
target data type requires bitshifts. Converting encoders and decoders
to use byte-level operations greatly improves the performance of these
functions, to the point that they can be competitive with
hand-implemented implementations, as our evaluation in
\autoref{sec:CaseStudies} will show. In order to do so without
compromising our correct-by-construction guarantee, we will justify
this conversion using an equivalence between bit-aligned and
byte-aligned functions.

The signatures of the byte-aligned functions instantiate the target
type of their bit-aligned versions to a byte buffer of fixed length
\lstinline|n|:
\begin{lstlisting}
AlignEncodeM S (n : nat) BIGSIGMA := S -> ByteBuff n -> nat -> BIGSIGMA -> Option (ByteBuff n * nat * BIGSIGMA)
AlignDecodeM S (n : nat) BIGSIGMA := ByteBuff n -> nat -> BIGSIGMA -> Option (S * nat * BIGSIGMA)
\end{lstlisting}
In addition to fixing the target type, byte-aligned encoders now take
the bytebuffer they write to, and both functions now carry the
index of the next byte to read/write. Both functions are
instances of the state and error monads, although we force
\lstinline|AlignDecodeM| to be read-only by threading the
byte buffer through the reader monad. We equip
\lstinline|AlignEncodeM| with a \lstinline|SetCurrentByte| operation
that sets the byte at the current index while updating that index, and
\lstinline|AlignDecodeM| with a corresponding
\lstinline|GetCurrentByte| operation for dequeuing bytes. We
define the twin equivalences used to justify the correctness of
byte-optimized functions as follows:
\begin{definition}[Correctness of Byte-Aligned Encoders]
  A byte-aligned encoder \lstinline|encode_bytes| and bit-aligned
  encoder \lstinline|encode_bits| are equivalent,
  $\mathsf{encode\_bits \simeq encode\_bytes}$, iff:
  \begin{itemize}
  \item \lstinline|encode_bytes| encodes the same bit sequence at the
    beginning of its byte buffer as \lstinline|encode_bits|.
  \item \lstinline|encode_bytes| fails when \lstinline|encode_bits|
    would write past the end of the fixed-length byte buffer.
  \item \lstinline|encode_bytes| fails whenever
    \lstinline|encode_bits| does.
  \end{itemize}
\end{definition}
\vspace{-.2cm}
\begin{definition}[Correctness of Byte-Aligned Decoders]
  A byte-aligned decoder \lstinline|decode_bytes| and bit-aligned
  decoder \lstinline|decode_bits| are equivalent,
  $\mathsf{decode\_bytes \backsimeq decode\_bytes}$, iff:
  \begin{itemize}
  \item \lstinline|decode_bytes| produces the same value as
    \lstinline|decode_bits|, while consuming the same number of bits.
  \item \lstinline|decode_bytes| fails when \lstinline|decode_bits|
    would write past the end of the fixed-length byte buffer.
  \item \lstinline|decode_bytes| fails whenever \lstinline|decode_bits| does.
  \end{itemize}
\end{definition}
\vspace{-.2cm} Armed with these definitions, we can build
transformation rules for deriving correct byte-aligned implementations
from bit-aligned functions in a similar manner to the previous
section.  
\autoref{fig:ByteAlignmentLemmas} gives examples of the rules for
byte-aligning decoders, the most important of these is the
\textsc{AlignDecSeq} which establishes that the byte-alignment
transformation can be decomposed through sequences. The rules for
byte-aligned encoders are similar. Note how \textsc{AlignDecByte}
proves an equivalence between dequeuing an 8-bit word and
\lstinline|AlignDecodeM|'s \lstinline|GetCurrentByte| operation. A key
part of automating derivations using these rules is associating
sequences of bit-aligned decoders so that this rule applies, as the
next section discusses in more detail.
\begin{figure}[!t]
  \vspace{-.1cm}

  \infrule[AlignDecSeq]
  {
    \mathsf{\scriptstyle decode\_bits_1~ \backsimeq~ decode\_bytes_1}
    \andalso
    \mathsf{\scriptstyle \forall v.\; decode\_bits_2\; v~ \backsimeq
      ~decode\_bytes_2 v}
  }
  {
    \mathsf{\scriptstyle decode\_bits_1 >\!\!>\!\!= decode\_bits_2~ \backsimeq~
      decode\_bytes_1 >\!\!>\!\!= decode\_bytes_2}
  }
  \begin{tabular}{ccc}
    \begin{minipage}{.3\textwidth}
    \infrule[AlignDecThrow]
    {
    }
    {
      \mathsf{\scriptstyle throw~ \backsimeq~ throw}
    }
  \end{minipage}
    &
    \begin{minipage}{.3\textwidth}
      \infrule[AlignDecByte]
      {
      }
      {
        \mathsf{\scriptstyle decode\_word_8 ~ \backsimeq~ GetCurrentByte}
      }
    \end{minipage}
    &
      \begin{minipage}{.3\textwidth}
        \infrule[AlignDecReturn]
        {
        }
        {
          \mathsf{\scriptstyle return\; a~ \backsimeq~ return\; a}
        }
      \end{minipage}
  \end{tabular}
\vspace{-.2cm}
\caption{A selection of byte-alignment rules for decoders. }
\label{fig:ByteAlignmentLemmas}
\vspace{-.35cm}

\end{figure}

\section{Automating Derivations}
\label{sec:synthesis}
As illustrated in \autoref{sec:narcissusTour}, \frameworkName provides
a set of tactics to help automate the derivations described above. The
tactic \verb|derive_encoder_decoder_pair| presented in that tour is
actually implemented via a pair of proof-automation tactics,
\lstinline|DeriveEncoder| and \lstinline|DeriveDecoder|, that derive
encoders and decoders, respectively.  Algorithm
\ref{alg:derive+decoder} presents the pseudocode algorithm for
\lstinline|DeriveDecoder|; \lstinline|DeriveEncoder| has a similar
implementation. In addition to the top-level format, $fmt$, this
tactic takes as input libraries of decoder-derivation and
byte-alignement rules, $drules$ and $arules$, which allow the tactic
to be extended to support new formats. \lstinline|DeriveDecoder| first
converts the input format to a normal form by right associating
sequences and collapsing nested applications of the format-composition
operator \lstinline|COMPOSER|. Next, the tactic attempts to derive a
bit-aligned decoder for $fmt$ via the \lstinline|ApplyRules|
subroutine that recursively applies the derivation rules in
$drules$. If a bit-level decoder is found, the algorithm again
normalizes the result using the monad laws and attempts to derive a
byte-aligned decoder by calling the \lstinline|AlignDecoder|
subroutine. Before diving into the details of the
\lstinline|ApplyRules| and \lstinline|AlignDecoder| tactics, we
emphasize that \lstinline|DeriveDecoder| is \emph{interactive}: if it
gets stuck on a goal it cannot solve with the current rule libraries,
it presents that goal to the user to solve interactively, as in the
derivation in \autoboxref{sensor4}{1}.

\begin{figure}[t]
  \begin{algorithm}[H]
  \caption{Derive a byte-aligned decoder from a format \label{alg:derive+decoder}}
\begin{tabular}{l|r}
  \begin{minipage}[t]{.52\linewidth}

  \begin{algorithmic}[1]
    \small
    \Function{\textsf{DeriveDecoder}}{$fmt, drules, arules$}

    \Statex \textbf{Input:} $fmt$: a format relation
    \Statex \hspace{\algorithmicindent} $drules$: set of decoder
    combinators derivation rules $arules$: set of byte-alignment
    transformation rules
    \Statex \textbf{Output:} $dec$: a byte-aligned decoder inverting $
    fmt$ \State $fmt_0 \leftarrow$ \Call{NormalizeFormat}{$fmt$}
    \State $dec \leftarrow$  \Call{ApplyRules}{$fmt_0, drules$}
    \State $dec_0 \leftarrow$ \Call{NormalizeDecoder}{$dec$} \State
    \Call{AlignDecoder}{$dec_0, arules$}
    \EndFunction

    \Statex
    \Function{\textsf{AlignDecoder}}{$dec, arules$}
    \For{$rule \leftarrow arules$}
    \Try
    \State $dec_0 \leftarrow$ \Call{DecAssoc}{$dec$}
    \State $\langle \overline{dec}, \mathcal{K}\rangle \leftarrow rule(dec)$
    \State $\overline{dec} \leftarrow$
    \Call{AlignDecoder}{$\overline{dec}, rules$}
    \State \Return $\mathcal{K}(\overline{dec})$
    \EndTry
    \EndFor

    \EndFunction

  \end{algorithmic}
\end{minipage}

  &
      \begin{minipage}[t]{.45\linewidth}
  \begin{algorithmic}[1]
    \small

    \Function{\textsf{ApplyRules}}{$fmt, drules$}

    \Try
    \State \Call{FinishDecoder}{$fmt$}
    \EndTry

    \For{$rule \leftarrow drules$}
    \Try
    \State $\langle \overline{fmt}, P, \mathcal{K}\rangle \leftarrow rule(fmt)$
    \State $\overline{dec} \leftarrow$
    \Call{ApplyRules}{$\overline{fmt}, rules$}
    \If {\Call{SolveSideConditions}{$P$}}
    \State \Return $\mathcal{K}(\overline{dec})$
    \EndIf
    \EndTry
    \EndFor
    \EndFunction

    \Statex
    \Statex

    \Function{\textsf{FinishDecoder}}{$fmt$}

    \Try
    \State $\langle \emptyset, (P_{src} P_{dec}), \mathcal{K}\rangle \leftarrow$ \textsc{DecDone}$(fmt)$
    \State $s \leftarrow$ \Call{ExtractView}{$P_{src}$}
    \State $b \leftarrow$ \Call{DecidePredicate}{$P_{dec}$}
    \State \Return \lstinline|if| $b$ \lstinline|then return| $s$ \lstinline|else fail|
    \EndTry
    \EndFunction

  \end{algorithmic}
\end{minipage}
\end{tabular}
\end{algorithm}
\vspace{-.5cm}
\end{figure}
\vspace{-.2cm}
\begin{align*}
  \mathcal{K} \in & \textsf{Cont} \triangleq \overline{\textsf{DecodeM S}_i~
                    \textsf{T}} \rightarrow \textsf{DecodeM S T} \\
  rule \in & \textsf{FormatM S T} \nrightarrow \langle
             \overline{\textsf{FormatM S}_i~\textsf{T}}, \textsf{Cont} \rangle
\end{align*}
In the implementation of \lstinline|ApplyRules|, derivation rules are
implemented as tactics that apply correctness lemmas to decompose the
current \lstinline|CorrectDecoder| goal into a set of simpler subgoals
in the standard interactive proof style. Conceptually,
\lstinline|ApplyRules| treats its derivation rules as partial
functions, each mapping a format to a triple of a (possibly empty) set of
subformats, a set of side conditions, and a continuation $\mathcal{K}$
that can construct a bit-aligned decoder from bit-aligned decoders for
those subformats, when those side conditions are satisfied. The
subformats represent the \roundtripAp premises of each derivation
rule, the side conditions capture its other premises, and the
continuation is the decoder in its conclusion. Thus,
\textsc{DecSeqProj} from \autoref{fig:DerivedCorrectDecoderLemmas} can
be thought of as a function that returns the subformats
$\mathsf{Q \cap fmt_1}$ and
$\mathsf{\{s~\mid~f~s=s' \land P \} \cap fmt_2}$, the side conditions
$\mathsf{\forall s.\; P\;s~ \rightarrow~ Q\; (f\; s)}$, and the
continuation $\lambda \mathsf{d_1 d_2. d_1 \lstinlines|++|_D~d_2}$,
when applied to a format of the form
$\mathsf{P \cap fmt_1 \circ f \lstinlines|++| fmt_2}$. A rule can fail
when the format in its conclusion does not match the current goal,
when its \lstinline|CorrectDecoder| subformats cannot be decoded, or
when its side conditions are not satisfied, e.g. \textsc{DecList}
fails when an appropriate length cannot
identified. \textsf{ApplyRules} first attempts to solve the
goal completely via the \textsf{FinishDecoder} tactic, which we will discuss
shortly.
\begin{wrapfigure}{r}{0.5\linewidth}
  \newcommand{\boxedDesc}[1]{%
    \fbox{\begin{minipage}{.8\linewidth}
        \centering
        \footnotesize\textit{#1}
      \end{minipage}}}
  \def\grayline{\tightrule[\color{Black}]{7pt}{5pt}{}}
  \centering
  \begin{ESHBlock}\input{listings/ltac0.v.esh.tex}\unskip\end{ESHBlock}
  \boxedDesc{Decompose (``destruct'') the source value,
    \lstinline|s|, with new variables for field values.}
  \begin{ESHBlock}\input{listings/ltac1.v.esh.tex}\unskip\end{ESHBlock}
  \boxedDesc{Substitute with equalities from source
    restriction (hypothesis \verb|H1|).}
  \begin{ESHBlock}\input{listings/ltac2.v.esh.tex}\unskip\end{ESHBlock}
  \boxedDesc{Variant of ``reflexivity'' solves the goal.}
  \caption{Example Ltac reconstruction of the original source value at
  the end of the derivation in \autoref{fig:SimpleDerivation}.}
  \label{fig:TermAutomation}
\end{wrapfigure}
If that tactic fails to find an appropriate decoder, the
algorithm iteratively attempts to apply the available rules to the
current format, starting with rules for base formats. If a rule is
successfully applied, the algorithm recursively calls
\textsf{ApplyRules} to derive decoders for any generated
subformats. If those derivations are successful, the algorithm applies
the continuation to the results and returns a finished decoder. If a
recursive call fails to process a subformat completely,
the tactic pauses and returns the corresponding \lstinline|CorrectDecoder|
subgoal, so that the user can see where the automation got
stuck. \textsf{AlignDecoder} is algorithmically similar to
\textsf{ApplyRules}, with the important modification that it attempts
to reassociate the topmost decoder using the \textsf{DecAssoc} tactic
before applying its transformation rules.

The \textsf{FinishDecoder} tactic warrants special
discussion. \textsf{FinishDecoder} attempts to finish a derivation of
a complete source value by finding instantiations of the
\lstinline|s| and \lstinline|b| metavariables in the \textsc{DecDone}
rule. Importantly, the original source value cannot be used for
either, but must instead be instantiated with values that \emph{only}
use previously parsed data. Automatically finding an instance of
$\mathsf{s}$ is particularly worrisome, as it is well-known that
Ltac, Coq's proof-automation language, does not provide good support
for introspecting into definitions of inductive types, and we would
like to use Ltac to construct records of fairly arbitrary types,
without relying on OCaml plugins. Thankfully, a combination of
standard tactics for case analysis and rewriting are up to the task.

Let us see how the \textsf{ExtractView} tactic attempts to discharge
the first proof obligation of \textsc{DecDone} for the derivation from
the previous section, which is presented in
\autoref{fig:TermAutomation}. This figure denotes the unknown
existential variable representing \lstinline|s| as
$\mathsf{\square_\mathsf{s}}$.  \textsf{ExtractView} first uses Coq's
standard \lstinline|destruct| tactic to perform case analysis on
\lstinline|s|, generating the second subgoal presented in
\autoref{fig:TermAutomation}, with occurrences of \lstinline|s|
replaced by its constructor applied to new variables
$\mathsf{x_1, x_2,}$ and $\mathsf{x_3}$. \textsf{ExtractView} then
attempts to remove any variables that are not in the scope of the
existential variable by rewriting the current goal using any
equalities about the original source value available in the
context. The resulting final goal equates
$\mathsf{\square_\mathsf{s}}$ to previously decoded values and can be
solved by unifying the two sides via the \lstinline|reflexivity|
tactic. Importantly, since $\mathsf{s}$ was not available when
$\mathsf{\square_\mathsf{s}}$ was quantified, this final tactic
\emph{only} succeeds when the rewritten term depends solely on
previously decoded data.  \textsf{FinishDecoder} then attempts to
solve a similar goal with a hole for \lstinline|b| using the
\textsf{DecidePredicate} tactic that employs known decision procedures
and simplifies away any tautologies, relying on a special typeclass to
resolve any user-defined predicates. We pause here to reemphasize
while \textsf{FinishDecoder} relies on heuristic-based proof
automation in a best effort attempt to solve the goal and is thus
incomplete, the failure of a tactic does not necessarily spell the end
of a derivation. By virtue of being implemented in an interactive
proof assistant, \frameworkName can loop users in when a derivation
gets stuck: if \textsf{FinishDecoder} cannot find a decoder for the
empty format, the user is presented a subgoal like the one at the top
of \autoref{fig:TermAutomation}, so that they can attempt to solve the
subgoal \emph{interactively}.

\section{Extending the Framework}
\label{sec:Extensions}

\label{sec:FrameworkExtensions}
As outlined in \autoref{sec:narcissusTour}, an extension to
\frameworkName consists of four pieces: a format, encoder and decoder
combinators, derivation rules, and automation for incorporating these
rules into \textsf{DeriveEncoder} and \textsf{DeriveDecoder}. As a
concrete example, consider the format of the Internet Protocol (IP)
checksum used in the IP headers, TCP segments, and UDP datagrams
featured in our case studies. \autoref{fig:IPChecksum} presents the
format, decoder combinator, and decoder derivation rule needed for
\frameworkName to support IP checksums. \lstinline{IP_Checksum_format}
is a higher-order combinator in the spirit of \lstinline{++}; the key
difference is that it uses the bitstrings produced by its subformat
parameters to build the IP checksum (the one's complement of the
bitstrings interpreted as lists of bytes), which it inserts between
the two encoded values to produce the output bitstring. The
\lstinline{IP_Checksum_decode} combinator has two subdecoder
parameters: it uses the first to calculate the number of bytes
included in the checksum, and then validates the checksum before
decoding the rest of the string using its second parameter.  The
derivation rule for this format guarantees that, when given the
correct number of bytes to include in the checksum, this test will
always succeed for uncorrupted data and that it can avoid parsing the
rest of the input otherwise. \autoref{fig:IPv4Example} presents a
complete example of this checksum combinator being used to derive
encoders and decoders for IP headers.

\begin{figure*}[!t]
  \hspace{-.2cm}
  \begin{varwidth}{\textwidth}
    \def\ESHInlineBlockFontSize{\small}
    \begin{ESHInlineBlock}[\ESHInlineBlockVerticalAlignment]\input{listings/ipchecksumfmt.v.esh.tex}\unskip\end{ESHInlineBlock}
    \vspace{\baselineskip}\tightrule[\color{LightGray}]{4pt}{6pt}{}
    \begin{ESHInlineBlock}[\ESHInlineBlockVerticalAlignment]\input{listings/ipchecksumdec.v.esh.tex}\unskip\end{ESHInlineBlock}
    \vspace{\baselineskip}\tightrule[\color{LightGray}]{4pt}{8pt}{}
    \centering
\begin{varwidth}{\textwidth}
  \infrule[DecChkSum]{
    \forall (s, \sigma_E, t, \sigma_E') \in \mathsf{fmt_1}\rightarrow
    \mathsf{length\; t} = \mathsf{len_1\;s}
    \andalso
    \forall s. \mathsf{len_1\;s~}\mathsf{mod}\; 8 = 0
    \\
    \forall (s, \sigma_E, t, \sigma_E') \in \mathsf{fmt_2} \rightarrow
    \mathsf{length\; t} = \mathsf{len_2\;s}
    \andalso
    \forall s. \mathsf{len_2\;s~}\mathsf{mod}\; 8 = 0
    \\
  \vspace{.2cm} 
  \mathsf{P} \cap \mathsf{fmt_1 ~+\hspace{-3pt}+~
    format\_unused\_word\;16 ~+\hspace{-3pt}+~ fmt_2} \; \roundtripAp
  \mathsf{dec_P} \\
  \mathsf{ P\,\cap\, subformat
    \roundtripPartMath{$\{(s, n) \mid len_1 s + 16 + len_2 s = n \times 8\}$}
    dec_M ~~ format_M}
  \\
  \mathsf{ subformat\; \le \; (fmt_1  ~+\hspace{-3pt}+~ format\_unused\_word 16
    ~+\hspace{-3pt}+~ fmt_2)}
}
{\mathsf{P} \cap
  \mathsf{IP\_Checksum\_format\; fmt_1 \; fmt_2\; } \; \roundtripAp \;
  \mathsf{IP\_Checksum\_decode\; dec_M\; dec_P} \; }
\end{varwidth}
  \end{varwidth}
  \vspace{-.3cm}
  \caption{Format, decoder, and decoder combinator for IP Checksums. }
\label{fig:IPChecksum}
\end{figure*}

\begin{figure}[!th]
  \def\ESHBlockFontSize{\small}
  \begin{ESHBlock}\input{listings/ipv4full.v.esh.tex}\unskip\end{ESHBlock}
  \vspace{-.1cm}
\caption{Format for IP version 4 headers, using the IP Checksum
  format.}
\label{fig:IPv4Example}
\end{figure}

\autoref{fig:IPv4Example} also includes a complete example of a new
decoder derivation tactic, which implements the \textsc{DecChkSum}
rule presented in \autoref{fig:IPChecksum}. \textsc{DecChkSum} is
similar to \textsc{DecSeqProj}, with a couple of key additional
assumptions. The first four of these ensure the bytestrings produced
by each subformat have constant length and are properly byte-aligned,
which is needed to prove the validity of the initial checksum
test. More interesting is the last assumption, which uses a decoder to
calculate the number of bytes to include in the checksum. As with the
\textsc{DecUnion} rule from \autoref{fig:CorrectDecoderLemmas}, this
framing allows \textsf{ApplyRules} to recursively discharge this
condition during a derivation. The
\lstinline{apply_new_combinator_rule} tactic applies this rule for IP
checksums, and attempts to discharge the first four assumptions by
using a database of facts about the lengths of encoded datatypes and
the modulus operator, relying on \textsf{ApplyRules} to derive
decoders for the subformats. Note that this tactic is a realization of
the logic of the body of \textsf{ApplyRules}'s loop, deriving
subdecoders recursively while discharging other subgoals
immediately. Other derivation rules included in \frameworkName have
similar implementations.

\section{Evaluation}
\label{sec:CaseStudies}

To evaluate the expressiveness and real-world applicability of \frameworkName,
we wrote specifications and derived implementations of encoders and decoders for
five of the most commonly used packet formats of the Internet protocol suite:
Ethernet, ARP, IPv4, TCP, UDP.  These formats were chosen to cover the full
TCP/IP stack while offering a wide variety of interesting features and challenges:

\begin{description}[leftmargin=.2cm,labelindent=0cm]
\item[\bf Checksums.] An IPv4 packet header contains a checksum equal to the
  one's-complement sum of the 16-bit words resulting from encoding all other
  fields of the header.
\item[\bf Pseudoheaders.] TCP and UDP segments also contain checksums, but they are
  computed on a segment's payload prefixed by a \emph{pseudoheader} that
  incorporates information from the IP layer.  This pseudoheader is not present
  in the encoded packet.
\item[\bf Unions.] An Ethernet frame header contains a 16-bit \texttt{EtherType} field,
  encoding either the length of the frame's payload (up to 1500 bytes) or a
  constant indicating which protocol the frame's payload encapsulates.  The two
  interpretations were originally conflicting, but the ambiguity was resolved in
  IEEE 802.3x-1997 by requiring all EtherType constants to be above 1535.
  This dichotomy is easily expressed in \frameworkName as a union format.
\item[\bf Constraints and underspecification.] TCP, UDP, and IP headers include
  underspecified or reserved-for-future-use bits, as well as fields with
  interdependencies (for example, the 16-bit urgent-pointer field of a TCP
  packet is only meaningful if its \texttt{URG} flag is set, and the options of a
  TCP packet mush be zero-padded to a 32-bit boundary equal to that specified
  by the packet's data-offset field).
\end{description}

The specifications of these formats are short and readable: new
formats typically requires 10 to 20 lines of declarative serialization
code and 10 to 20 lines of record-type, enumerated-type, and
numeric-constant declarations.  In addition to the base set of
formats, these specifications leverage a few TCP/IP-specific
extensions including checksums, pseudoheader checksums, and custom
index functions for union types.

The decoders that our framework produces are reasonably efficient and
sufficiently full-featured to be used as drop-in replacements for all
encoding and decoding components of a typical TCP/IP stack.  In the
rest of this section, we describe our extraction methodology and
support our claims by presenting performance benchmarks and reporting
on a fork of the native-OCaml \texttt{mirage-tcpip} library used in
the MirageOS unikernel, rewired to use our code to parse and decode
network packets.  We use Coq's extraction mechanism to obtain a
standalone OCaml library, using OCaml's integers to represent machine
words and natural numbers, a native-code checksum implementation, and
custom array data structures for the bytestrings and vectors that
encoders and decoders operate on. These custom data structures, as well as a
subset of the rewrite rules used during the final byte-alignment phase, are
unverified and thus part of our trusted base.

\subsection{Benchmarking}

\begin{wrapfigure}{r}{0.6\linewidth}
  \includegraphics[width=\linewidth]{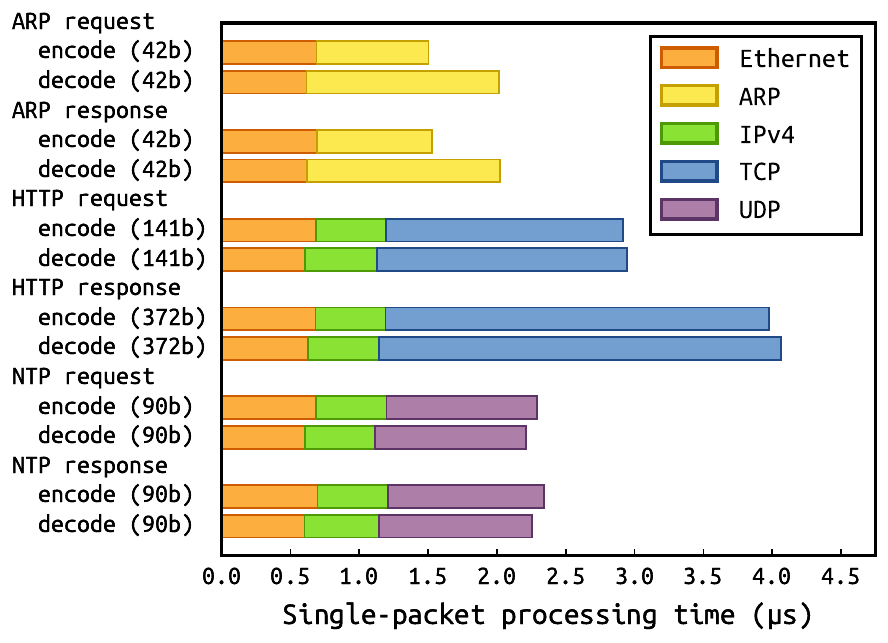}
  \caption{Processing times for various network packets on an Intel Core i7-4810MQ CPU @ 2.80GHz.  Each row shows how each layer of the network stack contributes to encoding and decoding times.  TCP and UDP checksums are computed over the entirety of the packet, payload included, which explains the higher processing times.  The HTTP and ARP payloads are a \texttt{GET} request to \texttt{http://nytimes.com} and a clock-synchronization request to \texttt{time.nist.gov}.}
  \label{fig:benchmarks}
\end{wrapfigure}

\autoref{fig:benchmarks} shows single-packet encoding and decoding times,
estimated by linearly regressing over the time needed to run batches of $n$ packet
serializations or deserializations for increasingly large values of $n$ (complete
experimental data, including 95\% confidence intervals, are provided as supplementary
material; they were obtained using the \texttt{Core\_bench} OCaml library~\cite{CoreBench}).

\subsection{Mirage OS Integration}

\textbf{MirageOS}~\cite{MirageOS} is a ``\textit{library operating
  system that constructs unikernels for secure, high-performance
  network applications}'': a collection of OCaml libraries that can be
assembled into a standalone kernel running on top of the Xen
hypervisor.  Security is a core feature of MirageOS, making it a
natural target to demonstrate integration of our encoders and decoders.  Concretely, this
entails patching the
\texttt{mirage-tcpip}\footnote{\url{https://github.com/mirage/mirage-tcpip}}
library to replace its serializers and deserializers by our own and evaluating the
resulting code in a realistic network application.  We chose the \texttt{mirage.io} website
(\texttt{mirage-www} on OPAM), which shows
that the overhead of using our decoders in a real-life application is
very small.

\paragraph{Setup} After extracting the individual encoders and decoders to OCaml, we
reprogrammed the TCP, UDP, IPv4, ARPv4, and Ethernet modules
of the \texttt{mirage-tcpip} library to use our
code optionally, and we recompiled everything.  This whole
process went smoothly: Mirage's test suite did not reveal issues
with our proofs, though we did have to adjust or disable some of Mirage's tests (for example, one test expected packets with incorrect
checksums to parse successfully, but our decoders reject them).

We strove to integrate into \texttt{mirage-tcpip} with minimal code changes: the vast majority
of our changes affect the five files concerned with marshaling and
unmarshaling our supported formats.  This yields a good estimate of the amount
of modification required (roughly 15 to 30 lines of glue code for each format), but it leaves lots of optimization opportunities unexplored: we
incur significant costs doing extra work and lining up mismatched representations.  Additionally, because we are strict about rejecting nonconforming
packets, we perform new work that Mirage was not performing, such as computing
checksums at parsing time or validating consistency constraints (Mirage's packet
decoders are a combination of hand-written bounds checks and direct reads at
automatically computed offsets into the packets).

\paragraph{Benchmarking} To evaluate the performance of the resulting
application, we ran the \texttt{mirage-www} server atop our modified
\texttt{mirage-tcpip} and measured the time needed to load pages from the \texttt{mirage.io} website as we
replaced each component by its
verified counterpart (we repeated each measurement 250 times, using the \texttt{win\-dow.per\-for\-mance.tim\-ing} counters in Firefox to measure page load times).   The incremental overhead of our verified decoders and encoders is minimal, ranging from less than 1\% on small pages to 0.5-4\% on large pages, such as the \texttt{blog/} page of the MirageOS website (accessing it causes the client to fetch about 4.2 MB of data, obtained through 36 HTTP requests spread across 1040 TCP segments):

\noindent{\centering\includegraphics[width=0.75\linewidth]{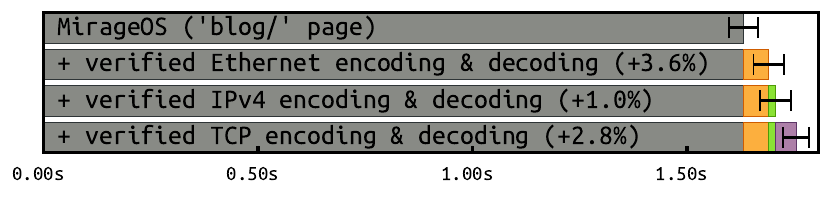}\par}


\section{Related Work}
\label{sec:RelatedWork}

\paragraph{Parsers for Context-Free Languages}
There is a long tradition of generating parsers for context-\emph{free}
languages from declarative Backus-Naur-form
specifications~\cite{YACC79,ANTLR1995} automatically.
Such generators may themselves have errors in them,
so in order to reduce the trusted code base of formally verified
compilers, there have been a number of efforts in verifying standalone
parsers for a variety of context-free languages~\cite{Barthwal+2009,
  Jourdan+2012, Koprowski+2011, Ridge+CFL, bernardy2016certified}. In
closely related work, the authors of RockSalt~\cite{Morrisett+2012}
developed a regular-expressions DSL, equipped with a relational
denotational semantics, in order to specify and generate verified
parsers from bitstrings into various instruction sets. In subsequent
work, \citet{Tan+2018} extended this DSL to support bidirectional
grammars in order to provide a uniform language for specifying and
generating both decoders and encoders, proving a similar notion of
consistency to what we present here. Importantly, all of these works
focus on languages that are insufficient for many network
protocols. Additionally, these parsers produce ASTs for types defined
by input grammars; these ASTs may need to be processed further
(possibly using \emph{semantic actions}) to recover original source
values. This processing phase must itself be verified to guarantee
correctness of the entire decoder.

\paragraph{Verification of Parsers for Network Protocol Formats}
A wide range of tools have been used to verify generated parsers for
binary network protocol formats~\cite{Collins+2017, Amin+17,
  Swamy+2016, Cheerios, lowstar}, including the SAW symbolic-analysis
engine, ~\cite{Dockins+2016}, the Frama-C analyzer~\cite{Cuoq+2012},
F*~\cite{Swamy+2016}, Agda~\cite{vanGeest}, and Coq. The correctness
properties of each project differ from \frameworkName's:
\citet{Amin+17} focus on memory safety. While \citet{lowstar} and
\citet{Collins+2017} prove that a pair of encoder and decoder
functions satisfy a round-trip property similar to ours, relying on
deterministic functions rules out many common formats, including DNS
packets, Google Protocol Buffers, or formats using ASN.1's BER
encoding. In addition, some of these approaches only support
constrained sets of formats: \citet{Collins+2017} are restricted to
ASN.1 formats, while \citet{Cheerios} requires the format to align
with the source type. \citet{vanGeest} also use a library of parsers
and pretty printers for a fixed set of data types to build
implementations, but they rely on verified datatype transformations to
support a more flexible set of formats, including IPv4 headers. More
closely related is the verified protocol-buffer compiler of
\citet{ProtobuffCPP}, whose development adopts \frameworkName's
definition of correctness for top-level decoders and reuses its format
for fixed-length words. That effort did not adopt \frameworkName's
combinator-based philosophy, as the compiler was hand-written and
manually verified in Coq, and is limited to formats in the fixed
data-description language of the Protocol Buffer standard.

\paragraph{Deductive Synthesis}
The idea of deriving correct-by-construction implementations from
specifications using deductive rules has existed for at least half a
century~\cite{EWD:EWD209, Manna+79}. Kestrel's
Specware~\cite{SpecWare} system was an seminal realization of this
idea, and has been used to implement correct-by-construction SAT
solvers~\cite{SpecWareSAT}, garbage collectors~\cite{SpecWareGC}, and
network protocols. Deductive approaches have been employed more
recently to interactively derive verified recursive functions in a
general purpose programming language~\cite{Leon} and cache-efficient
implementations of divide-and-conquer
algorithms~\cite{Bellmania}. Very closely related is the Fiat
framework~\cite{FiatPOPL15, FiatSNAPL} for interactively deriving
abstract data types inside of Coq with domain-specific specifications
and proof automation. \frameworkName builds upon Fiat, reusing its
implementation of the aforementioned nondeterminism monad, in addition
to some of its datatype definitions and common datatype definitions
and general proof automation tactics. These dependencies represent a
small portion of the Fiat library; the remaining aspects of
\frameworkName presented in this paper are novel, including the
problem domain, the specifications of decoder/encoder correctness and
the formulation of formats used in those specifications, and the
derivation tactics for encoders and decoders.

\paragraph{Parser-Combinator Libraries}
There is a long history in the functional-programming community of
using combinators~\cite{leijen+2001} to eliminate the burden of
writing parsers by hand, but less attention has been paid to the
question of how to generate both encoders and
decoders. Kennedy~\cite{Kennedy+2004} presents a library of
combinators that package serializers and deserializers for data types
to/from bytestrings (these functions are also called picklers and
unpicklers) into a common typeclass. A similar project extended
Haskell's Arrow class~\cite{Hughes+2000} with a reverse arrow in order
to represent invertible functions~\cite{Alimarine+2005}. In more
closely related work, Rendel and Ostermann developed a combinator
library for writing pairs of what they term partial
isomorphisms~\cite{Rendel+2010}; that is, partial functions each of
which correctly invert all values in the other's range. The authors
give a denotational semantics for their EDSL using a relational
interpretation that closely mirrors \frameworkName's. Importantly,
proofs of correctness for these libraries, where they exist, are
strictly informal.

\paragraph{Bidirectional / Invertible Programming Languages}
Mu et. al present a functional language in which only injective
functions can be defined, allowing users to invert every program
automatically~\cite{Mu+2004}. The authors give a relational semantics
to this language, although every program in the language is a
function. The authors show how to embed noninjective programs in their
language automatically by augmenting them with sufficient information
to invert each computation. They prove that this additional
information can be dropped given a user-provided inversion function.

Boomerang~\cite{bohannon+2008} is a bidirectional programming language
for projecting transformations on a data view back to the original
source data; in contrast to \frameworkName, Boomerang does not require
that the original source values can be recovered from a target
view. Boomerang programs are built using a collection of lens
combinators, which include get, put, and create operations for
transporting modifications between source and target representations.
While Boomerang originally synthesized functions that assumed that
every source value had a canonical target representation, it has since
been extended with quotient lenses that relax this
restriction~\cite{foster+2008}. The recently developed Optician
tool~\cite{miltner+2017} synthesizes Boomerang programs that implement
bijective string transformations from regular expressions describing
source and target formats and sets of user-provided disambiguating
examples. The format-decoding problem differs from the lens setting in
that lenses consider how to recover a new source value from an
updated target value given full knowledge of the \emph{old} source
value, while decoding must work given only a single target value.

\paragraph{Extensible Format-Description Languages}
Interface generators like XDR~\cite{XDR}, ASN.1~\cite{ASN2001}, Apache
Avro~\cite{Avro}, and Protocol Buffers~\cite{protobuf} generate
encoders and decoders from user-defined data schemes. The underlying
data format for these frameworks can be context-sensitive, but this
format is defined by the system, however, preventing data exchange
between programs using different frameworks. The lack of fine-grained
control over the target representation prevents users from extending
the format, which could bring benefits in dimensions like compactness,
even ignoring the need for compatibility with widely used standards.

The binpac compiler~\cite{Binpac2006} supports a
data-format-specification language specifically developed for network
protocols but does not support extending the language beyond the
built-in constructs. More recent frameworks, like
PADS~\cite{PADS2005}, PacketTypes~\cite{PacketTypes}, and
Datascript~\cite{Datascript}, feature sophisticated data-description
languages with support for complex data dependencies and constraints
for specific data schemes but also lack support for extensions.
Nail~\cite{Nail} is a tool for synthesizing parsers and generators
from formats in a high-level declarative language. Nail unifies the
data-description format and internal data layout into a single
specification and allows users to specify and automatically check
dependencies between encoded fields. More importantly, Nail natively
supports extensions to its parsers and generators via user-defined
\emph{stream transformations} on the encoded data, allowing it to
capture protocol features that other frameworks cannot. However, Nail
provides no formal guarantees, and these transformations can introduce
bugs violating the framework's safety properties.  We also note two
other differences between Nail and \frameworkName.  First, Nail has
many more orthogonal primitives than \frameworkName, as our primitives
may be considered to be little more than the definition of decoder
correctness.  Second, while Nail provides flexibility in describing
binary formats, it maps each format to a fixed C struct type, where
\frameworkName is compatible with arbitrary Coq types.

\section{Conclusion}

We have presented \frameworkName, a framework for
specifying and deriving correct-by-construction encoders and decoders
for non-context-free formats in the style of parser-combinator
libraries. This framework provides fine-grained control over the shape
of encoded data, is extensible with user-defined formats and
implementation strategies, has a small set of core definitions
augmented with a library of common formats, and produces
machine-checked proofs of soundness for derived decoders and
encoders. We evaluated the expressiveness of \frameworkName by
deriving decoders and encoders for several standardized formats and
demonstrated the utility of the derived functions by incorporating
them into the OCaml-based Mirage operating system.
\bibliography{narcissus}



\end{document}

%% file: Preamble.tex
\definecolor{Black}{rgb}{0.1,0.1,0.1}
\definecolor{White}{rgb}{1,1,1}
\definecolor{LightGray}{rgb}{0.8,0.8,0.8}
\definecolor{DarkGray}{rgb}{0.2,0.2,0.2}
\definecolor{MediumGray}{rgb}{0.8,0.8,0.8}
\definecolor{LightRed}{rgb}{0.8,0.4,0.4}
\definecolor{LightBlue}{rgb}{0.4,0.8,0.4}
\definecolor{LightGreen}{rgb}{0.6,0.6,0.8}
\definecolor{LightPurple}{RGB}{179,40,244}
\definecolor{DPurple}{RGB}{79,21,95}
\definecolor{DGreen}{RGB}{8,116,41}
\definecolor{DOrange}{RGB}{145,20,20}
\definecolor{DBlue}{RGB}{0,79,121}
\definecolor{DYellow}{RGB}{145,126,0}
\definecolor{HPurple}{RGB}{119,61,135}
\definecolor{HGreen}{RGB}{48,156,81}
\definecolor{HOrange}{RGB}{185,60,60}
\definecolor{HBlue}{RGB}{15,119,161}
\definecolor{HYellow}{RGB}{185,166,0}

\definecolor{StringIDs}{rgb}{0.5,0.5,0.5}

\newcommand{\neghalfthinspace}{\kern -0.0833em}
\newcommand\SubScript[1]{\ensuremath{_\textsf{#1}}}

\usepackage{xspace}
\newcommand\frameworkName{\textsc{Narcissus}\xspace}

\def\lstinlines{\lstinline[basicstyle=\footnotesize]}

\renewcommand*\descriptionlabel[1]{\hspace\labelsep\normalfont #1}

\newenvironment{smallaligned*}{\nobreak\footnotesize\noindent \csname aligned*\endcsname}{\csname endaligned*\endcsname}

\newcommand{\roundtrip}{\hspace{-.2cm}$\begin{tikzpicture}[baseline=-0.63ex, ->, thick]%
\node (A) { };%
\node (B) [right of=A, node distance=3em] { };%

\path[every node/.style={font=\sffamily\small}]
    (A) edge[bend left=20] node [right] {} (B)
    (B) edge[bend left=20] node [left] {} (A);
\end{tikzpicture}$}

\newcommand{\roundtripAp}{\hspace{-.2cm}$\begin{tikzpicture}[baseline=-0.63ex, ->, thick]%
\node (A) { };%
\node (B) [right of=A, node distance=3em] { };%
\node (C) [right of=A, node distance=1.5em] {$\approx$ };%

\path[every node/.style={font=\sffamily\small}]
    (A) edge[bend left=20] node [right] {} (B)
    (B) edge[bend left=20] node [left] {} (A);
\end{tikzpicture}$}

\newcommand{\RoundTripMathAp}{\hspace{-.1cm}\begin{tikzpicture}[baseline=-0.63ex, ->, thick]%
\node (A) { };%
\node (B) [right of=A, node distance=3em] { };%
\node (C) [right of=A, node distance=1.5em] {$\approx$ };%
\path[every node/.style={font=\sffamily\small}]
    (A) edge[bend left=20] node [right] {} (B)
    (B) edge[bend left=20] node [left] {} (A);
\end{tikzpicture}\hspace{-.1cm}}

\newcommand{\roundtripPart}[1]{\hspace{-.1cm}$\begin{tikzpicture}[baseline=-0.63ex,
    ->, thin, color=DarkGray!, dashed]%
\node (A) { };%
\node (B) [right of=A, node distance=3em] { };%
\node (C) [right of=A, node distance=1.5em] {$\approx$ };%
\node (D) [below of=B, node distance=.5em] {\textcolor{DarkGray}{\tiny #1}};%

\path[every node/.style={font=\sffamily\small}]
    (A) edge[bend left=20] node [right] {} (B)
    (B) edge[bend left=20] node [left] {} (A);
  \end{tikzpicture}$\hspace{-.1cm}}

\newcommand{\roundtripPartMath}[1]{\hspace{-.1cm}\begin{tikzpicture}[baseline=-0.63ex,
    ->, thin, color=DarkGray!, dashed]%
\node (A) { };%
\node (B) [right of=A, node distance=3em] { };%
\node (C) [right of=A, node distance=1.5em] {$\approx$ };%
\node (D) [below of=B, node distance=.5em] {\textcolor{DarkGray}{\tiny #1}};%

\path[every node/.style={font=\sffamily\small}]
    (A) edge[bend left=20] node [right] {} (B)
    (B) edge[bend left=20] node [left] {} (A);
\end{tikzpicture}\hspace{-.1cm}}

\newcommand{\fmapMath}{\hspace{-.05cm}
\begin{tikzpicture}[baseline=-0.63ex]
\node [diamond, draw, inner sep=0pt, scale=.7] {\textsf{\$}};%
\end{tikzpicture}
}

\input{esh-preamble}
\input{narcissus.esh-pv}

\newcommand*{\parunskipped}{\par\nopagebreak\vspace{-\parskip}}

\newcommand{\tightrule}[4][]{%
  \parunskipped\nopagebreak\vspace{-\baselineskip}\nopagebreak\vspace{#2}
  \noindent{#1\hrulefill#4}\parunskipped\vspace{#3}\nointerlineskip}

\newcounter{EncDecExample}
\newcommand{\boxref}[3][]{\hyperref[esh:#2.#3]{#1\ref{esh:#2}.#3}}
\newcommand{\autoboxref}[2]{\boxref[box~]{#1}{#2}}
\newcommand{\autoboxedexampleref}[1]{\hyperref[esh:#1]{Example~\ref{esh:#1}}}

\newcommand{\codeBlockHeader}[1]{%
  \tightrule{5pt}{4pt}{%
    \raisebox{\fboxrule-\height}[0pt][0pt]{%
      \setlength{\fboxsep}{2pt}%
      \fbox{\scriptsize\arabic{EncDecExample}.#1}}}}

\newcommand{\spacedCodeBlock}[2][]{{%
  \parunskipped\vspace{0.5\abovedisplayskip}
  #1{\small#2}
  \parunskipped\vspace{0.5\belowdisplayskip}}}

\newcommand{\NarcissusExample}[1]{{%
    \setlength{\ESHSkip}{0pt}
    \refstepcounter{EncDecExample}\label{esh:#1}
    \label{esh:#1.1}\codeBlockHeader{1: User input}
    \begin{ESHBlock}\input{listings/#1def.v.esh.tex}\unskip\end{ESHBlock}
    \label{esh:#1.2}\codeBlockHeader{2: Encoder}
    \begin{ESHBlock}\input{listings/#1enc.v.esh.tex}\unskip\end{ESHBlock}
    \label{esh:#1.3}\codeBlockHeader{3: Decoder}
    \begin{ESHBlock}\input{listings/#1dec.v.esh.tex}\unskip\end{ESHBlock}}}

\newlength{\NarcissusMargin}
\newenvironment{NarcissusNarrow}{%
  \list{}{\rightmargin\NarcissusMargin\leftmargin\NarcissusMargin}\item[]
}{%
  \endlist
}

\newcommand{\NarcissusExampleOneCol}[1]{{%
    \renewcommand{\baselinestretch}{0.95}
    \spacedCodeBlock{%
      \begin{NarcissusNarrow}
        \NarcissusExample{#1}
      \end{NarcissusNarrow}}}}

\newcommand{\NarcissusPartialExample}[1]{{%
    \renewcommand{\baselinestretch}{0.95}
    \setlength{\ESHSkip}{0pt}
    \refstepcounter{EncDecExample}\label{esh:#1}
    \spacedCodeBlock{%
      \begin{NarcissusNarrow}
      \label{esh:#1.1}\codeBlockHeader{1: User input}
      \begin{ESHBlock}\input{listings/#1def.v.esh.tex}\unskip\end{ESHBlock}
      \end{NarcissusNarrow}}}}

\definecolor{LightGray}{HTML}{DDDDDD}
\newlength{\graylineabove}
\newlength{\graylinebelow}
\newcommand{\grayline}{\tightrule[\color{LightGray}]{\graylineabove}{\graylinebelow}{}}

\def\ESHBlockFontSize{\small}
\renewcommand{\ESHBlockHook}{%
  \ESHBlockFontSize
  \def\ESHEol{\ifvmode\else\nopagebreak[3]\newline\fi}
  \def\ESHEmptyLine{\vspace{-\baselineskip}\grayline}}
\renewcommand{\ESHInlineHook}{%
  \renewcommand{\_}{\-\textunderscore\-}\hyphenpenalty=50}
\def\ESHInlineBlockEol{\cr}
\def\ESHInlineBlockFontSize{\small}
\renewcommand{\ESHInlineBlockHook}{%
  \ESHInlineBlockFontSize
  \def\ESHEol{\ESHInlineBlockEol}%
  \def\ESHEmptyLine{%
    \gdef\ESHInlineBlockEol{\gdef\ESHInlineBlockEol{\cr}}%
    {\hspace{-1pt}\color{LightGray}\leaders\hbox{\rule[\graylinebelow]{1pt}{0.4pt}}\hfill}%
    \\[\dimexpr\graylineabove+\graylinebelow-\ESHBaselineskip\relax]}}
\usepackage[T1]{fontenc}

\usepackage{ifthen}
\let\OldClassWarning\ClassWarning
\renewcommand{\ClassWarning}[2]{\ifthenelse{\equal{#1}{acmart}}{}{\OldClassWarning{#1}{#2}}}

\mdfdefinestyle{MyFrame}{%
    linecolor=black,
    outerlinewidth=2pt,
    innertopmargin=4pt,
    innerbottommargin=4pt,
    innerrightmargin=4pt,
    innerleftmargin=4pt,
        leftmargin = 4pt,
        rightmargin = 4pt
        }

%% file: narcissus.esh-pv.tex
\ESHpvDefineVerb{verb}|H1|{\ESHInline{\ESHBol{}H\ESHRaise{-0.25}{1}}}
\ESHpvDefineVerb{verb}|derive_encoder_decoder_pair|{\ESHInline{\ESHBol{}\textcolor[HTML]{00008B}{derive\_encoder\_decoder\_pair}}}
\ESHpvDefineVerb{verb}|S|{\ESHInline{\ESHBol{}S}}
\ESHpvDefineVerb{verb}|format_reading|{\ESHInline{\ESHBol{}\textcolor[HTML]{75507B}{format\_reading}}}
\ESHpvDefineVerb{verb}|reading|{\ESHInline{\ESHBol{}reading}}
\ESHpvDefineVerb{verb}|format_nat 8 ◦ length|{\ESHInline{\ESHBol{}\textcolor[HTML]{75507B}{format\_nat}\ESHSpace{}8\ESHSpace{}\ESHUnicodeSubstitution{\ensuremath{\circ}}\ESHSpace{}length}}
\ESHpvDefineVerb{verb}|data|{\ESHInline{\ESHBol{}data}}
\ESHpvDefineVerb{verb}|format_list|{\ESHInline{\ESHBol{}\textcolor[HTML]{75507B}{format\_list}}}
\ESHpvDefineVerb{verb}|"HUMIDITY"|{\ESHInline{\ESHBol{}\textcolor[HTML]{5C3566}{"HUMIDITY"}}}
\ESHpvDefineVerb{verb}|0b01|{\ESHInline{\ESHBol{}0b01}}
\ESHpvDefineVerb{verb}|"TEMP"|{\ESHInline{\ESHBol{}\textcolor[HTML]{5C3566}{"TEMP"}}}
\ESHpvDefineVerb{verb}|0b00|{\ESHInline{\ESHBol{}0b00}}
\ESHpvDefineVerb{verb}|format_enum|{\ESHInline{\ESHBol{}\textcolor[HTML]{75507B}{format\_enum}}}
\ESHpvDefineVerb{verb}|format_const|{\ESHInline{\ESHBol{}\textcolor[HTML]{75507B}{format\_const}}}
\ESHpvDefineVerb{verb}|kind|{\ESHInline{\ESHBol{}kind}}
\ESHpvDefineVerb{verb}|format_unused_word|{\ESHInline{\ESHBol{}\textcolor[HTML]{75507B}{format\_unused\_word}}}
\ESHpvDefineVerb{verb}|0x00|{\ESHInline{\ESHBol{}0x00}}
\ESHpvDefineVerb{verb}|stationID|{\ESHInline{\ESHBol{}stationID}}
\ESHpvDefineVerb{verb}|⋅|{\ESHInline{\ESHBol{}\ESHUnicodeSubstitution{\ESHMathSymbol{\cdot}}}}
\ESHpvDefineVerb{verb}|low_bits|{\ESHInline{\ESHBol{}low\_bits}}
\ESHpvDefineVerb{verb}|high_bits|{\ESHInline{\ESHBol{}high\_bits}}
\ESHpvDefineVerb{verb}|SetCurrentByte|{\ESHInline{\ESHBol{}\textcolor[HTML]{75507B}{SetCurrentByte}}}
\ESHpvDefineVerb{verb}|b ∘ a|{\ESHInline{\ESHBol{}b\ESHSpace{}\ESHUnicodeSubstitution{\ESHMathSymbol{\circ}}\ESHSpace{}a}}
\ESHpvDefineVerb{verb}|a ⋙ b|{\ESHInline{\ESHBol{}a\ESHSpace{}\ESHUnicodeSubstitution{\ESHMathSymbol{\ggg}}\ESHSpace{}b}}
\ESHpvDefineVerb{verb}|⋙|{\ESHInline{\ESHBol{}\ESHUnicodeSubstitution{\ESHMathSymbol{\ggg}}}}
\ESHpvDefineVerb{verb}|GetCurrentByte|{\ESHInline{\ESHBol{}\textcolor[HTML]{75507B}{GetCurrentByte}}}
\ESHpvDefineVerb{verb}|≫|{\ESHInline{\ESHBol{}\ESHUnicodeSubstitution{\ESHMathSymbol{\gg}}}}
\ESHpvDefineVerb{verb}|<-|{\ESHInline{\ESHBol{}\ESHUnicodeSubstitution{\ESHMathSymbol{\leftarrow}}}}
\ESHpvDefineVerb{verb}|None|{\ESHInline{\ESHBol{}None}}
\ESHpvDefineVerb{verb}|invariant|{\ESHInline{\ESHBol{}invariant}}
\ESHpvDefineVerb{verb}|++|{\ESHInline{\ESHBol{}++}}
\ESHpvDefineVerb{verb}|format_word|{\ESHInline{\ESHBol{}\textcolor[HTML]{75507B}{format\_word}}}
\ESHpvDefineVerb{verb}|format|{\ESHInline{\ESHBol{}format}}
\ESHpvDefineVerb{verb}|Record|{\ESHInline{\ESHBol{}\textcolor[HTML]{346604}{Record}}}
\ESHpvDefineVerb{verb}|sensor_msg|{\ESHInline{\ESHBol{}sensor\_msg}}
\DeclareRobustCommand*{\verb}{\ESHpvLookupVerb{verb}}

%% file: listings/sensor3fail.v.esh.tex
\ESHBol{}CorrectDecoder\ESHSpace{}(\textcolor[HTML]{75507B}{format\_list}\ESHSpace{}\textcolor[HTML]{75507B}{format\_word}\ESHSpace{}\ESHUnicodeSubstitution{\ensuremath{\circ}}\ESHSpace{}data\ESHSpace{}++\ESHSpace{}\ESHUnicodeSubstitution{\ESHMathSymbol{\ldots}})\ESHSpace{}\textcolor[HTML]{B35000}{?d}

%% file: listings/sensor3fail2.v.esh.tex
\ESHBol{}\textcolor[HTML]{204A87}{\ESHUnicodeSubstitution{\ESHMathSymbol{\forall}}}\ESHSpace{}\textcolor[HTML]{B35000}{msg}:\ESHSpace{}sensor\_msg,\ESHSpace{}stationID\ESHSpace{}msg\ESHSpace{}=\ESHSpace{}sid\ESHSpace{}\ESHUnicodeSubstitution{\ESHMathSymbol{\rightarrow}}\ESHSpace{}length\ESHSpace{}msg.(measurements)\ESHSpace{}=\ESHSpace{}\textcolor[HTML]{B35000}{?}\textcolor[HTML]{346604}{Goal}

%% file: listings/sensor4fail.v.esh.tex
\ESHBol{}CorrectDecoder\ESHSpace{}(\textcolor[HTML]{75507B}{format\_nat}\ESHSpace{}8\ESHSpace{}\ESHUnicodeSubstitution{\ensuremath{\circ}}\ESHSpace{}length\ESHSpace{}\ESHUnicodeSubstitution{\ensuremath{\circ}}\ESHSpace{}data\ESHSpace{}++\ESHSpace{}\ESHUnicodeSubstitution{\ESHMathSymbol{\ldots}})\ESHSpace{}\textcolor[HTML]{B35000}{?d}

%% file: listings/sensor4fail2.v.esh.tex
\ESHBol{}\textcolor[HTML]{204A87}{\ESHUnicodeSubstitution{\ESHMathSymbol{\forall}}}\ESHSpace{}\textcolor[HTML]{B35000}{data}:\ESHSpace{}sensor\_msg,\ESHSpace{}invariant\ESHSpace{}data\ESHSpace{}\ESHUnicodeSubstitution{\ESHMathSymbol{\wedge}}\ESHSpace{}stationID\ESHSpace{}data\ESHSpace{}=\ESHSpace{}proj\ESHSpace{}\ESHUnicodeSubstitution{\ESHMathSymbol{\rightarrow}}\ESHEol
\ESHBol{}\ESHSpace{}\ESHSpace{}\ESHSpace{}\ESHSpace{}\ESHSpace{}\ESHSpace{}\ESHSpace{}\ESHSpace{}\ESHSpace{}\ESHSpace{}\ESHSpace{}\ESHSpace{}\ESHSpace{}\ESHSpace{}\ESHSpace{}\ESHSpace{}\ESHSpace{}\ESHSpace{}\ESHSpace{}\ESHSpace{}length\ESHSpace{}data.(measurements)\ESHSpace{}{<}\ESHSpace{}2\ESHRaise{0.50}{16}

%% file: listings/ltac0.v.esh.tex
\ESHBol{}H\ESHRaise{-0.25}{1}\ESHSpace{}:\ESHSpace{}length\ESHSpace{}s.data\ESHSpace{}{<}\ESHSpace{}2\ESHRaise{0.50}{8}\ESHSpace{}\ESHUnicodeSubstitution{\ESHMathSymbol{\wedge}}\ESHSpace{}s.stationId\ESHSpace{}=\ESHSpace{}w\ESHEol
\ESHBol{}\ESHSpace{}\ESHSpace{}\ESHSpace{}\ESHUnicodeSubstitution{\ESHMathSymbol{\wedge}}\ESHSpace{}length\ESHSpace{}s.data\ESHSpace{}=\ESHSpace{}ln\ESHSpace{}\ESHUnicodeSubstitution{\ESHMathSymbol{\wedge}}\ESHSpace{}s.data\ESHSpace{}=\ESHSpace{}l\ESHEol
\ESHBol{}\ESHEmptyLine{}\ESHEol
\ESHBol{}\ESHUnicodeSubstitution{\scalebox{0.75}{\ensuremath{\square}}}\ESHRaise{-0.25}{s}\ESHSpace{}=\ESHSpace{}s

%% file: listings/ltac1.v.esh.tex
\ESHBol{}H\ESHRaise{-0.25}{1}\ESHSpace{}:\ESHSpace{}length\ESHSpace{}s.data\ESHSpace{}{<}\ESHSpace{}2\ESHRaise{0.50}{8}\ESHSpace{}\ESHUnicodeSubstitution{\ESHMathSymbol{\wedge}}\ESHSpace{}x\ESHRaise{-0.25}{1}\ESHSpace{}=\ESHSpace{}w\ESHEol
\ESHBol{}\ESHSpace{}\ESHSpace{}\ESHSpace{}\ESHUnicodeSubstitution{\ESHMathSymbol{\wedge}}\ESHSpace{}length\ESHSpace{}x\ESHRaise{-0.25}{2}\ESHSpace{}=\ESHSpace{}ln\ESHSpace{}\ESHUnicodeSubstitution{\ESHMathSymbol{\wedge}}\ESHSpace{}x\ESHRaise{-0.25}{2}\ESHSpace{}=\ESHSpace{}l\ESHEol
\ESHBol{}\ESHEmptyLine{}\ESHEol
\ESHBol{}\ESHUnicodeSubstitution{\scalebox{0.75}{\ensuremath{\square}}}\ESHRaise{-0.25}{s}\ESHSpace{}=\ESHSpace{}\{\ESHSpace{}stationId\ESHSpace{}:=\ESHSpace{}x\ESHRaise{-0.25}{1};\ESHSpace{}data\ESHSpace{}:=\ESHSpace{}x\ESHRaise{-0.25}{2}\ESHSpace{}\}

%% file: listings/ltac2.v.esh.tex
\ESHBol{}H\ESHRaise{-0.25}{1}\ESHSpace{}:\ESHSpace{}length\ESHSpace{}s.data\ESHSpace{}{<}\ESHSpace{}2\ESHRaise{0.50}{8}\ESHEol
\ESHBol{}\ESHEmptyLine{}\ESHEol
\ESHBol{}\ESHUnicodeSubstitution{\scalebox{0.75}{\ensuremath{\square}}}\ESHRaise{-0.25}{s}\ESHSpace{}=\ESHSpace{}\{\ESHSpace{}stationId\ESHSpace{}:=\ESHSpace{}w;\ESHSpace{}data\ESHSpace{}:=\ESHSpace{}l\ESHSpace{}\}

%% file: listings/ipchecksumfmt.v.esh.tex
\ESHBol{}\textcolor[HTML]{346604}{Let}\ESHSpace{}\textcolor[HTML]{A40000}{IP\_Checksum\_format}\ESHSpace{}\{S\}\ESHSpace{}format\ESHRaise{-0.25}{1}\ESHSpace{}format\ESHRaise{-0.25}{2}\ESHSpace{}(\textcolor[HTML]{B35000}{s}\ESHSpace{}:\ESHSpace{}S)\ESHSpace{}:=\ESHSpace{}\textcolor[HTML]{204A87}{\ESHUnicodeSubstitution{\ESHMathSymbol{\lambda}}}\ESHSpace{}\textcolor[HTML]{B35000}{ctx}\ESHSpace{}\ESHUnicodeSubstitution{\ESHMathSymbol{\Rightarrow}}\ESHEol
\ESHBol{}\ESHSpace{}\ESHSpace{}{`}(p,\ESHSpace{}ctx)\ESHSpace{}\ESHUnicodeSubstitution{\ESHMathSymbol{\leftarrow}}\ESHSpace{}format\ESHRaise{-0.25}{1}\ESHSpace{}s\ESHSpace{}ctx;\ESHEol
\ESHBol{}\ESHSpace{}\ESHSpace{}{`}(q,\ESHSpace{}ctx)\ESHSpace{}\ESHUnicodeSubstitution{\ESHMathSymbol{\leftarrow}}\ESHSpace{}format\ESHRaise{-0.25}{2}\ESHSpace{}s\ESHSpace{}(addE\ESHSpace{}ctx\ESHSpace{}16);\ESHEol
\ESHBol{}\ESHSpace{}\ESHSpace{}c\ESHSpace{}\ESHUnicodeSubstitution{\ESHMathSymbol{\leftarrow}}\ESHSpace{}\{\ESHSpace{}\textcolor[HTML]{B35000}{c}\ESHSpace{}:\ESHSpace{}word\ESHSpace{}16\ESHSpace{}|\ESHSpace{}\textcolor[HTML]{204A87}{\ESHUnicodeSubstitution{\ESHMathSymbol{\forall}}}\ESHSpace{}\textcolor[HTML]{B35000}{ext},\ESHEol
\ESHBol{}\ESHSpace{}\ESHSpace{}\ESHSpace{}\ESHSpace{}\ESHSpace{}\ESHSpace{}\ESHSpace{}\ESHSpace{}IPChecksum\_Valid\ESHSpace{}(bin\_measure\ESHSpace{}(p\ESHSpace{}++\ESHSpace{}(encode\_word\ESHSpace{}c)\ESHSpace{}++\ESHSpace{}q))\ESHEol
\ESHBol{}\ESHSpace{}\ESHSpace{}\ESHSpace{}\ESHSpace{}\ESHSpace{}\ESHSpace{}\ESHSpace{}\ESHSpace{}\ESHSpace{}\ESHSpace{}\ESHSpace{}\ESHSpace{}\ESHSpace{}\ESHSpace{}\ESHSpace{}\ESHSpace{}\ESHSpace{}\ESHSpace{}\ESHSpace{}\ESHSpace{}\ESHSpace{}\ESHSpace{}\ESHSpace{}\ESHSpace{}\ESHSpace{}(p\ESHSpace{}++\ESHSpace{}(encode\_word\ESHSpace{}c)\ESHSpace{}++\ESHSpace{}q\ESHSpace{}++\ESHSpace{}ext)\ESHSpace{}\};\ESHEol
\ESHBol{}\ESHSpace{}\ESHSpace{}\textcolor[HTML]{346604}{ret}\ESHSpace{}(p\ESHSpace{}++\ESHSpace{}(encode\_word\ESHSpace{}c)\ESHSpace{}++\ESHSpace{}q,\ESHSpace{}ctx).

%% file: listings/ipchecksumdec.v.esh.tex
\ESHBol{}\textcolor[HTML]{346604}{Let}\ESHSpace{}\textcolor[HTML]{A40000}{IP\_Checksum\_decode}\ESHSpace{}(\textcolor[HTML]{B35000}{bin}\ESHSpace{}:\ESHSpace{}B)\ESHSpace{}(\textcolor[HTML]{B35000}{env}\ESHSpace{}:\ESHSpace{}CacheDecode)\ESHSpace{}:=\ESHEol
\ESHBol{}\ESHSpace{}\ESHSpace{}{`}(n,\ESHSpace{}\_,\ESHSpace{}\_)\ESHSpace{}\ESHUnicodeSubstitution{\ESHMathSymbol{\leftarrow}}\ESHSpace{}decode\_measure\ESHSpace{}bin\ESHSpace{}env;\ESHEol
\ESHBol{}\ESHSpace{}\ESHSpace{}\textcolor[HTML]{204A87}{if}\ESHSpace{}checksum\_Valid\_dec\ESHSpace{}(n\ESHSpace{}*\ESHSpace{}8)\ESHSpace{}bin\ESHSpace{}\textcolor[HTML]{204A87}{then}\ESHSpace{}decodeA\ESHSpace{}bin\ESHSpace{}env\ESHEol
\ESHBol{}\ESHSpace{}\ESHSpace{}\textcolor[HTML]{204A87}{else}\ESHSpace{}None

%% file: listings/ipv4full.v.esh.tex
\ESHBol{}\textcolor[HTML]{346604}{Record}\ESHSpace{}\textcolor[HTML]{A40000}{IPv4\_Packet}\ESHSpace{}:=\ESHEol
\ESHBol{}\ESHSpace{}\ESHSpace{}\{\ESHSpace{}TotalLength:\ESHSpace{}word\ESHSpace{}16;\ESHSpace{}ID:\ESHSpace{}word\ESHSpace{}16;\ESHEol
\ESHBol{}\ESHSpace{}\ESHSpace{}\ESHSpace{}\ESHSpace{}DF:\ESHSpace{}\ESHUnicodeSubstitution{\ESHMathSymbol{\mathbb{B}}};\ESHSpace{}MF:\ESHSpace{}\ESHUnicodeSubstitution{\ESHMathSymbol{\mathbb{B}}};\ESHSpace{}FragmentOffset:\ESHSpace{}word\ESHSpace{}13;\ESHSpace{}TTL:\ESHSpace{}word\ESHSpace{}8;\ESHEol
\ESHBol{}\ESHSpace{}\ESHSpace{}\ESHSpace{}\ESHSpace{}Protocol:\ESHSpace{}EnumType\ESHSpace{}[\textcolor[HTML]{5C3566}{"ICMP"};\ESHSpace{}\textcolor[HTML]{5C3566}{"TCP"};\ESHSpace{}\textcolor[HTML]{5C3566}{"UDP"}];\ESHEol
\ESHBol{}\ESHSpace{}\ESHSpace{}\ESHSpace{}\ESHSpace{}SourceAddress:\ESHSpace{}word\ESHSpace{}32;\ESHSpace{}DestAddress:\ESHSpace{}word\ESHSpace{}32;\ESHEol
\ESHBol{}\ESHSpace{}\ESHSpace{}\ESHSpace{}\ESHSpace{}Options:\ESHSpace{}list\ESHSpace{}(word\ESHSpace{}32)\ESHSpace{}\}.\ESHEol
\ESHBol{}\ESHEmptyLine{}\ESHEol
\ESHBol{}\textcolor[HTML]{346604}{Definition}\ESHSpace{}\textcolor[HTML]{A40000}{ProtocolTypeCodes}\ESHSpace{}:=\ESHSpace{}\ESHSlantItalic{\textcolor[HTML]{5F615C}{(*\ESHSpace{}Protocol\ESHSpace{}Numbers\ESHSpace{}from\ESHSpace{}[RFC\ESHRaise{-0.25}{5237}]\ESHSpace{}*)}}\ESHEol
\ESHBol{}\ESHSpace{}\ESHSpace{}[0\ESHWeightBold{\textcolor[HTML]{204A87}{b}}00000001\ESHSpace{}\ESHSlantItalic{\textcolor[HTML]{5F615C}{(*\ESHSpace{}ICMP:\ESHSpace{}1\ESHSpace{}*)}};\ESHSpace{}0\ESHWeightBold{\textcolor[HTML]{204A87}{b}}00000110\ESHSpace{}\ESHSlantItalic{\textcolor[HTML]{5F615C}{(*\ESHSpace{}TCP:\ESHSpace{}6\ESHSpace{}*)}};\ESHSpace{}0\ESHWeightBold{\textcolor[HTML]{204A87}{b}}00010001\ESHSpace{}\ESHSlantItalic{\textcolor[HTML]{5F615C}{(*\ESHSpace{}UDP:\ESHSpace{}17\ESHSpace{}*)}}].\ESHEol
\ESHBol{}\ESHEmptyLine{}\ESHEol
\ESHBol{}\textcolor[HTML]{346604}{Definition}\ESHSpace{}\textcolor[HTML]{A40000}{IPv4\_Packet\_Format}\ESHSpace{}:\ESHSpace{}FormatM\ESHSpace{}IPv4\_Packet\ESHSpace{}ByteString\ESHSpace{}:=\ESHEol
\ESHBol{}\ESHSpace{}\ESHSpace{}\ESHSpace{}\ESHSpace{}(\textcolor[HTML]{75507B}{format\_nat}\ESHSpace{}4\ESHSpace{}\ESHUnicodeSubstitution{\ensuremath{\circ}}\ESHSpace{}(constant\ESHSpace{}4)\ESHEol
\ESHBol{}\ESHSpace{}\ESHSpace{}++\ESHSpace{}\textcolor[HTML]{75507B}{format\_nat}\ESHSpace{}4\ESHSpace{}\ESHUnicodeSubstitution{\ensuremath{\circ}}\ESHSpace{}(plus\ESHSpace{}5)\ESHSpace{}\ESHUnicodeSubstitution{\ensuremath{\circ}}\ESHSpace{}\ESHWeightBold{\textcolor[HTML]{204A87}{@}}length\ESHSpace{}\_\ESHSpace{}\ESHUnicodeSubstitution{\ensuremath{\circ}}\ESHSpace{}Options\ESHEol
\ESHBol{}\ESHSpace{}\ESHSpace{}++\ESHSpace{}\textcolor[HTML]{75507B}{format\_unused\_word}\ESHSpace{}8\ESHSpace{}\ESHSlantItalic{\textcolor[HTML]{5F615C}{(*\ESHSpace{}TOS\ESHSpace{}Field!\ESHSpace{}*)}}\ESHEol
\ESHBol{}\ESHSpace{}\ESHSpace{}++\ESHSpace{}\textcolor[HTML]{75507B}{format\_word}\ESHSpace{}\ESHUnicodeSubstitution{\ensuremath{\circ}}\ESHSpace{}TotalLength\ESHEol
\ESHBol{}\ESHSpace{}\ESHSpace{}++\ESHSpace{}\textcolor[HTML]{75507B}{format\_word}\ESHSpace{}\ESHUnicodeSubstitution{\ensuremath{\circ}}\ESHSpace{}ID\ESHEol
\ESHBol{}\ESHSpace{}\ESHSpace{}++\ESHSpace{}\textcolor[HTML]{75507B}{format\_unused\_word}\ESHSpace{}1\ESHSpace{}\ESHSlantItalic{\textcolor[HTML]{5F615C}{(*\ESHSpace{}Unused\ESHSpace{}flag!\ESHSpace{}*)}}\ESHEol
\ESHBol{}\ESHSpace{}\ESHSpace{}++\ESHSpace{}\textcolor[HTML]{75507B}{format\_bool}\ESHSpace{}\ESHUnicodeSubstitution{\ensuremath{\circ}}\ESHSpace{}DF\ESHEol
\ESHBol{}\ESHSpace{}\ESHSpace{}++\ESHSpace{}\textcolor[HTML]{75507B}{format\_bool}\ESHSpace{}\ESHUnicodeSubstitution{\ensuremath{\circ}}\ESHSpace{}MF\ESHEol
\ESHBol{}\ESHSpace{}\ESHSpace{}++\ESHSpace{}\textcolor[HTML]{75507B}{format\_word}\ESHSpace{}\ESHUnicodeSubstitution{\ensuremath{\circ}}\ESHSpace{}FragmentOffset\ESHEol
\ESHBol{}\ESHSpace{}\ESHSpace{}++\ESHSpace{}\textcolor[HTML]{75507B}{format\_word}\ESHSpace{}\ESHUnicodeSubstitution{\ensuremath{\circ}}\ESHSpace{}TTL\ESHEol
\ESHBol{}\ESHSpace{}\ESHSpace{}++\ESHSpace{}\textcolor[HTML]{75507B}{format\_enum}\ESHSpace{}ProtocolTypeCodes\ESHSpace{}\ESHUnicodeSubstitution{\ensuremath{\circ}}\ESHSpace{}Protocol)\ESHEol
\ESHBol{}\ESHSpace{}\ESHSpace{}\textcolor[HTML]{75507B}{ThenChecksum}\ESHSpace{}IPChecksum\_Valid\ESHSpace{}\textcolor[HTML]{75507B}{OfSize}\ESHSpace{}16\ESHSpace{}\textcolor[HTML]{75507B}{ThenCarryOn}\ESHEol
\ESHBol{}\ESHSpace{}\ESHSpace{}\ESHSpace{}\ESHSpace{}(\textcolor[HTML]{75507B}{format\_word}\ESHSpace{}\ESHUnicodeSubstitution{\ensuremath{\circ}}\ESHSpace{}SourceAddress\ESHEol
\ESHBol{}\ESHSpace{}\ESHSpace{}++\ESHSpace{}\textcolor[HTML]{75507B}{format\_word}\ESHSpace{}\ESHUnicodeSubstitution{\ensuremath{\circ}}\ESHSpace{}DestAddress\ESHEol
\ESHBol{}\ESHSpace{}\ESHSpace{}++\ESHSpace{}\textcolor[HTML]{75507B}{format\_list}\ESHSpace{}\textcolor[HTML]{75507B}{format\_word}\ESHSpace{}\ESHUnicodeSubstitution{\ensuremath{\circ}}\ESHSpace{}Options).\ESHEol
\ESHBol{}\ESHEmptyLine{}\ESHEol
\ESHBol{}\textcolor[HTML]{346604}{Definition}\ESHSpace{}\textcolor[HTML]{A40000}{IPv4\_Packet\_OK}\ESHSpace{}(\textcolor[HTML]{B35000}{ipv\ESHRaise{-0.25}{4}}\ESHSpace{}:\ESHSpace{}IPv4\_Packet)\ESHSpace{}:=\ESHEol
\ESHBol{}\ESHSpace{}\ESHSpace{}(length\ESHSpace{}ipv\ESHRaise{-0.25}{4}.(Options))\ESHSpace{}{<}\ESHSpace{}11\ESHSpace{}\ESHUnicodeSubstitution{\ESHMathSymbol{\wedge}}\ESHEol
\ESHBol{}\ESHSpace{}\ESHSpace{}20\ESHSpace{}+\ESHSpace{}4\ESHSpace{}*\ESHSpace{}(length\ESHSpace{}ipv\ESHRaise{-0.25}{4}.(Options))\ESHSpace{}{<}\ESHSpace{}wordToNat\ESHSpace{}ipv\ESHRaise{-0.25}{4}.(TotalLength).\ESHEol
\ESHBol{}\ESHEmptyLine{}\ESHEol
\ESHBol{}\textcolor[HTML]{346604}{Ltac}\ESHSpace{}\textcolor[HTML]{A40000}{new\_encoder\_rules}\ESHSpace{}::=\ESHEol
\ESHBol{}\ESHSpace{}\ESHSpace{}\textcolor[HTML]{204A87}{match}\ESHSpace{}goal\ESHSpace{}\textcolor[HTML]{204A87}{with}\ESHEol
\ESHBol{}\ESHSpace{}\ESHSpace{}\ESHSpace{}\ESHSpace{}\ESHUnicodeSubstitution{\ESHMathSymbol{\vdash}}\ESHSpace{}CorrectAlignedEncoder\ESHSpace{}(\_\ESHSpace{}\textcolor[HTML]{75507B}{ThenChecksum}\ESHSpace{}\_\ESHSpace{}\textcolor[HTML]{75507B}{OfSize}\ESHSpace{}\_\ESHSpace{}\textcolor[HTML]{75507B}{ThenCarryOn}\ESHSpace{}\_)\ESHSpace{}\_\ESHSpace{}\ESHUnicodeSubstitution{\ESHMathSymbol{\Rightarrow}}\ESHEol
\ESHBol{}\ESHSpace{}\ESHSpace{}\ESHSpace{}\ESHSpace{}\textcolor[HTML]{00008B}{eapply}\ESHSpace{}\ESHWeightBold{\textcolor[HTML]{204A87}{@}}CorrectAlignedEncoderForIPChecksumThenC\ESHEol
\ESHBol{}\ESHSpace{}\ESHSpace{}\textcolor[HTML]{204A87}{end}.\ESHEol
\ESHBol{}\ESHEmptyLine{}\ESHEol
\ESHBol{}\textcolor[HTML]{346604}{Ltac}\ESHSpace{}\textcolor[HTML]{A40000}{apply\_new\_combinator\_rule}\ESHSpace{}::=\ESHEol
\ESHBol{}\ESHSpace{}\ESHSpace{}\textcolor[HTML]{204A87}{match}\ESHSpace{}goal\ESHSpace{}\textcolor[HTML]{204A87}{with}\ESHEol
\ESHBol{}\ESHSpace{}\ESHSpace{}|\ESHSpace{}H\ESHSpace{}:\ESHSpace{}cache\_inv\_Property\ESHSpace{}\textcolor[HTML]{B35000}{?mnd}\ESHSpace{}\_\ESHEol
\ESHBol{}\ESHSpace{}\ESHSpace{}\ESHSpace{}\ESHSpace{}\ESHUnicodeSubstitution{\ESHMathSymbol{\vdash}}\ESHSpace{}CorrectDecoder\ESHSpace{}\_\ESHSpace{}\_\ESHSpace{}\_\ESHSpace{}\_\ESHSpace{}(\textcolor[HTML]{B35000}{?fmt\ESHRaise{-0.25}{1}}\ESHSpace{}\textcolor[HTML]{75507B}{ThenChecksum}\ESHSpace{}\_\ESHSpace{}\textcolor[HTML]{75507B}{OfSize}\ESHSpace{}\_\ESHSpace{}\textcolor[HTML]{75507B}{ThenCarryOn}\ESHSpace{}\textcolor[HTML]{B35000}{?fmt\ESHRaise{-0.25}{2}})\ESHSpace{}\_\ESHSpace{}\_\ESHSpace{}\_\ESHSpace{}\ESHUnicodeSubstitution{\ESHMathSymbol{\Rightarrow}}\ESHEol
\ESHBol{}\ESHSpace{}\ESHSpace{}\ESHSpace{}\ESHSpace{}\textcolor[HTML]{00008B}{eapply}\ESHSpace{}compose\_IPChecksum\_format\_correct{'}\ESHSpace{}\textcolor[HTML]{204A87}{with}\ESHSpace{}(format\ESHRaise{-0.25}{1}\ESHSpace{}:=\ESHSpace{}fmt\ESHRaise{-0.25}{1});\ESHEol
\ESHBol{}\ESHSpace{}\ESHSpace{}\ESHSpace{}\ESHSpace{}[\ESHSpace{}\textcolor[HTML]{FF0000}{exact}\ESHSpace{}H\ESHSpace{}|\ESHSpace{}\textcolor[HTML]{B452CD}{repeat}\ESHSpace{}calculate\_length\_ByteString\ESHSpace{}|\ESHSpace{}\textcolor[HTML]{B452CD}{repeat}\ESHSpace{}calculate\_length\_ByteString\ESHEol
\ESHBol{}\ESHSpace{}\ESHSpace{}\ESHSpace{}\ESHSpace{}|\ESHSpace{}solve\_mod\_8\ESHSpace{}|\ESHSpace{}solve\_mod\_8\ESHSpace{}|\ESHSpace{}\textcolor[HTML]{00008B}{intros};\ESHSpace{}normalize\_format;\ESHSpace{}apply\_rules\ESHEol
\ESHBol{}\ESHSpace{}\ESHSpace{}\ESHSpace{}\ESHSpace{}|\ESHSpace{}normalize\_format;\ESHSpace{}apply\_rules\ESHSpace{}|\ESHSpace{}solve\_Prefix\_Format\ESHSpace{}]\ESHEol
\ESHBol{}\textcolor[HTML]{204A87}{end}.\ESHEol
\ESHBol{}\ESHEmptyLine{}\ESHEol
\ESHBol{}\textcolor[HTML]{346604}{Let}\ESHSpace{}\textcolor[HTML]{A40000}{enc\_dec}\ESHSpace{}:\ESHSpace{}\textcolor[HTML]{75507B}{EncoderDecoderPair}\ESHSpace{}IPv4\_Packet\_Format\ESHSpace{}IPv4\_Packet\_OK.\ESHEol
\ESHBol{}\textcolor[HTML]{346604}{Proof}.\ESHSpace{}\textcolor[HTML]{00008B}{derive\_encoder\_decoder\_pair}.\ESHSpace{}\textcolor[HTML]{346604}{Defined}.\ESHEol
\ESHBol{}\ESHEmptyLine{}\ESHEol
\ESHBol{}\textcolor[HTML]{346604}{Let}\ESHSpace{}\textcolor[HTML]{A40000}{IPv4\_encoder}\ESHSpace{}:=\ESHSpace{}encoder\_impl\ESHSpace{}enc\_dec.\ESHEol
\ESHBol{}\textcolor[HTML]{346604}{Let}\ESHSpace{}\textcolor[HTML]{A40000}{IPv4\_decoder}\ESHSpace{}:=\ESHSpace{}decoder\_impl\ESHSpace{}enc\_dec.